\documentclass[final,3p,times,authoryear]{elsarticle}
\DeclareRobustCommand{\disambiguate}[3]{#2~#3}

\usepackage{amssymb}

\usepackage{caption}
\usepackage{subcaption}
\usepackage{graphicx}
\usepackage[dvips]{epsfig}
\usepackage{xtab}
\usepackage{booktabs}
\usepackage{lscape}
\usepackage{rotating}
\usepackage{fixltx2e}
\usepackage{natbib} 
\usepackage{import} 
\usepackage[usenames, dvipsnames]{color}
\usepackage{float}
\usepackage[utf8]{inputenc}
\usepackage{afterpage}
\usepackage{lineno}
\usepackage{csquotes}
\usepackage{siunitx} 
\usepackage{multirow}

\journal{Icarus}

\begin{document}

\begin{frontmatter}


\title{The Elemental Abundances (with Uncertainties) of the Most Earth-like Planet}

\author[label1,label2]{Haiyang S. Wang\corref{cor1}}
\ead{haiyang.wang@anu.edu.au}
\author[label1,label2,label3]{Charles H. Lineweaver}
\author[label2,label3]{Trevor R. Ireland}

\address[label1]{Research School of Astronomy and Astrophysics, The Australian National University, Canberra, ACT 2611, Australia}
\address[label2]{Planetary Science Institute, The Australian National University, Canberra, ACT 2611, Australia}
\address[label3]{Research School of Earth Sciences, The Australian National University, Canberra, ACT 2601, Australia}
\cortext[cor1]{Corresponding author. Mailing address: Mt Stromlo Observatory (ANU Research School of Astronomy and Astrophysics), Cotter Road, Weston Creek, ACT 2611, Australia}

\begin{abstract}
	
To first order, the Earth as well as other rocky planets in the Solar System and rocky exoplanets orbiting other stars, are 
refractory
pieces of the stellar nebula out of which they formed.
To estimate the chemical composition of rocky exoplanets based on their stellar hosts' elemental abundances, we need a better understanding of the devolatilization that produced the Earth.
To quantify the chemical relationships between the Earth, the Sun and other bodies in the Solar System, the elemental abundances of the bulk Earth are required. The key to comparing Earth's composition with those of other objects is to have a determination of the bulk composition with an appropriate estimate of uncertainties. Here we present concordance estimates (\textit {with uncertainties}) of the elemental abundances  of the bulk Earth, which can be used in such studies. First we compile, combine and renormalize a large set of heterogeneous literature values of the primitive mantle (PM) and of the core. We then integrate standard radial density profiles of the Earth and renormalize them to the current best estimate for the mass of the Earth. Using estimates of the uncertainties in i) the density profiles, ii) the core-mantle boundary and iii) the inner core boundary, we employ standard error propagation to obtain a core mass fraction of $32.5 \pm 0.3$ wt\%. Our bulk Earth abundances are the weighted sum of our concordance core abundances and concordance PM abundances. 
Unlike previous efforts, the uncertainty on the core mass fraction is propagated to the uncertainties on the bulk Earth elemental abundances.
Our concordance estimates for the abundances of Mg, Sn, Br, B, Cd and Be are significantly lower than previous estimates of the bulk Earth. Our concordance estimates for the abundances of Na, K, Cl, Zn, Sr, F, Ga, Rb, Nb, Gd, Ta, He, Ar, and Kr are significantly higher. The uncertainties on our elemental abundances usefully calibrate the unresolved discrepancies between standard Earth models under various geochemical and geophysical assumptions. 
\end{abstract}


\end{frontmatter}

\section{Introduction}
\label{intro}
The number of known rocky exoplanets is rapidly increasing. 
Transit photometry and radial velocity measurements, when combined, yield rough estimates of the densities and therefore
mineralogies of these exoplanets.
Independent and potentially more precise estimates of the chemical composition of these rocky planets can be made based on the known elemental abundances of their host stars combined with estimates of the devolatilization process that produced the rocky planets from their stellar nebulae.
To proceed with this strategy, we need to quantify the devolatilization that produced the Earth from the solar nebula.
Knowledge of the bulk elemental abundances of the Earth with uncertainties is an important part of this research. 
The elemental abundances of bulk Earth (including both the bulk silicate Earth and the core) can tell us a more complete story of the potentially universal accretion and fractionation processes that produce rocky planets from nebular gas during star formation. Uncertainties associated with the bulk Earth composition are needed to compare and quantify compositional differences between the Earth, Sun, and other solar system bodies.
Such comparisons can lead to a more detailed understanding of devolatilization and the chemical relationship between a terrestrial planet and its host star. The bulk Earth elemental abundances will help determine what mixture of meteorites, comets and other material produced the Earth \citep{Drake2002, Burbine2004} and can also help determine the width of the feeding zone of the Earth in the protoplanetary disk \citep{Chambers2001, Kaib2015}. 

There are many stages of compositional fractionation between the collapse of a stellar nebula, the evolution of a proto-planetary disk and a final rocky planet. Composition- and position-dependent differences in the duration and strength of the various fractionating processes can lead to a variety of outcomes. The different (but somewhat similar) compositions of Earth, Mars and Vesta are a measure of these variations within our own Solar System.

A major challenge to estimating the bulk chemical composition of the Earth is that we can only sample the upper part of the possibly heterogeneous mantle, and we have no direct access to its deep interior, and even less to the core \citep{Allegre2001}. 
Early studies on primitive mantle (PM) elemental abundances include \cite{ONeill1991}, \cite{Kargel1993}, \cite{McDonough1995}, and \cite{ONeill1998}. Bulk elemental abundances with uncertainties \citep{Allegre2001} were reported 16 years ago but much work on PM abundances and on core abundances (usually separately) has been done since then \citep[e.g.,][]{Lyubetskaya2007, Palme2014b, Rubie2011, Wood2013, Hirose2013}.
The determination of uncertainties is central to the quantification of elemental abundances but has been a missing priority in previous work. The most recent, highly cited estimates of the elemental abundances of the bulk Earth do not include uncertainties \citep{McDonough2003, McDonough2008}. 

Elemental abundance discrepancies are in large part model-dependent \citep{McDonough2016}, but over the past 15 years progress has been made in making
more plausible models.
Our knowledge of the core (and therefore of the bulk Earth) has increased significantly: high pressure experiments yield improved partition coefficients of siderophiles \citep{Siebert2013} and improved affinities of light elements for iron \citep{Ricolleau2011, Mookherjee2011}. Seismic velocites through the core provide increasingly precise constraints on densities and on mineral physics models \citep{Vocadlo2007, Li2014, Badro2014}. Better subduction models \citep{Poitrasson2015} and estimates of the degree of homogeneity of the mantle \citep{Javoy2010, Nakajima2015a, McDonough2016} provide new constraints that are being included in the upper and lower mantle abundance estimates. Better observations of geo-neutrinos \citep{Bellini2010, Gando2011, Bellini2013, Gando2013} provide new thermal constraints for the abundances of heat-producing elements in the Earth \citep[e.g.][]{Sramek2013, Huang2013}. 
The large majority of the literature on elemental abundances of the Earth, involving either the analysis of the PM or of the core, are increasingly important and when combined, yield improved elemental abundances of the bulk Earth composition and more realistic uncertainties. 

Our main research goal is to analyze the compositional differences between the Earth, Sun and other solar system bodies and from this comparison quantify the devolatilization of stellar material that leads to rocky planets.
This requires estimates of the bulk Earth abundances {\it with uncertainties}. These bulk abundances and their uncertainties are poorly constrained and often ignored in the literature. Motivated by this, we make a concordance estimate of the bulk Earth elemental abundances and their uncertainties. The words \textquote{concordance estimate} specifically mean a compositional estimate that is representative of, and concordant with previous estimates. The aim of this work therefore is not to resolve the discrepancies between competing models and assumptions but to construct a concordance model ({\it with uncertainties}) that represents current knowledge of bulk Earth composition and calibrates unresolved discrepancies. 
We envisage that if the discrepancies can be resolved, a better formulation for estimating uncertainties might be forthcoming. However, at this stage, some of the arguments concerning the derivation of models, or values resulting from these models, appear intractable. Nevertheless, there is an essential need for uncertainties in the estimates if we are to make progress in comparing planetary objects.

We organize this paper as follows. In Sect. 2 we present the concordance estimate for the composition of PM.  In Sect. 3 we present the concordance estimate of the core; in Sect. 4 we make a new estimate (with uncertainty) of the core mass fraction of the Earth, using it as the weighting factor to combine the PM with the core to yield our concordance estimates for the bulk Earth. In Sect. 5 we discuss details of how our work differs from previous work and some unresolved issues that might affect our results. 

\section{Composition of the Primitive Mantle}
\label{sec:pm}
\subsection{Data sources}
\label{sec:pm-1}

Earth's primitive mantle (PM) or bulk silicate Earth (BSE) is the mantle existing after core segregation but before the extraction of continental and oceanic crust and the degassing of volatiles \citep[][]{Sun1982, Kargel1993, Saal2002, McDonough2003, Lyubetskaya2007, Palme2014b}.
There are two major, and partially-overlapping modeling strategies for estimating the PM composition.

The peridotite model is based on the analysis of chemical data from basalts and periodotite massifs.
Peridotite-basalt melting trends yield an estimate of the PM composition
\citep[e.g.,][]{Ringwood1979, Sun1982, McDonough1995, McDonough2003, Lyubetskaya2007}.
The peridotite model has a number of intrinsic problems, including the non-uniqueness of melting trends, large scatter in the data from
mid-ocean ridge basalts (MORB) and from ocean island basalts (OIB), and the difficulty of imposing multiple cosmochemical constraints on refractory lithophile element (RLE) abundances, often resulting in model-dependent, poorly quantified uncertainties \citep{Lyubetskaya2007}. 
 
The cosmochemical model is based on the identification of Earth with a particular class of chondritic or achondritic meteorites or their mixtures \citep[e.g.,][]{Morgan1980, Javoy2010, Fitoussi2016}, along with a number of assumptions on accretion and fractionation processes \citep{Allegre2001}. 
The cosmochemical model uses chondritic ratios of RLEs and volatility trends \citep[e.g.,][]{Wanke1988, Palme2003, Palme2014b}.
\cite{Palme2003} and \cite{Palme2014b} present a core-mantle mass balance approach for calculating the primitive mantle composition. 
This approach requires an accurate determination of magnesium number (Mg \#  = molar Mg/(Mg+Fe)).
A reasonable range for Mg\# can be inferred from fertile mantle periodites.

Our aim is not to resolve the differences between these strategies and models but to construct a concordance model whose mean values and uncertainties adequately represent current knowledge and disagreement.  

 \subsection{Concordance PM Estimate}
 \label{sec:pm-2}
 
 We construct our concordance PM abundances from three major papers reporting PM abundances \citep{Lyubetskaya2007,McDonough2008,Palme2014b}, supplemented with noble gas abundances from \cite{Marty2012} and \cite{Halliday2013}.
 \cite{McDonough2008} is an updated version of their pioneering peridotite model \citep{McDonough1995,McDonough2003}. Updates of some abundances, for example
  W, K and Pb, can be found in \cite{Arevalo2008} and \cite{Arevalo2009}. 
 \cite{Lyubetskaya2007} performed a principal component analysis of the same peridodite database but with different model parameters. 
 \cite{Palme2014b} is largely based on a cosmochemical model using mass balance and is an updated version of their
pioneering earlier work \citep{Palme2003}.
 
 Challenges to combining these three data sets are:
 
 i) \cite{McDonough2008} report no uncertainties, however the PM abundance uncertainties reported in \cite{McDonough1995} approximately reflect current uncertainties. Thus, in our analysis, we attach them to the PM abundances reported in \cite{McDonough2008}.
 
 ii) \cite{Lyubetskaya2007} do not report abundances of four volatiles; C, H, N, O.
 
 iii) \cite{Palme2014b} do not report uncertainties for C, N, Se, Te, In, Hg and Bi, while \cite{Lyubetskaya2007} do not report uncertainties for Te, In, Hg, Bi, Ag, Cd and Tl. We assign the \cite{McDonough1995} uncertainties to these elements. 
 
 iv) none of these three PM models include noble gas abundances. In a preliminary attempt to be more comprehensive, we supplement these three PM models with recent noble gas abundances: the atmospheric model of \cite{Marty2012} and three different models (layered mantle, impact erosion, and basaltic glass) from \cite{Halliday2013}.
This range encompasses recent results from \cite{Dauphas2014} \citep{Marty2016}.
See Appendix A.1 for details.
 
\begin{figure}[H]
	\begin{center}
		\includegraphics[trim=0.9cm 2.4cm 0.8cm 3.0cm, scale=0.75,angle=0]{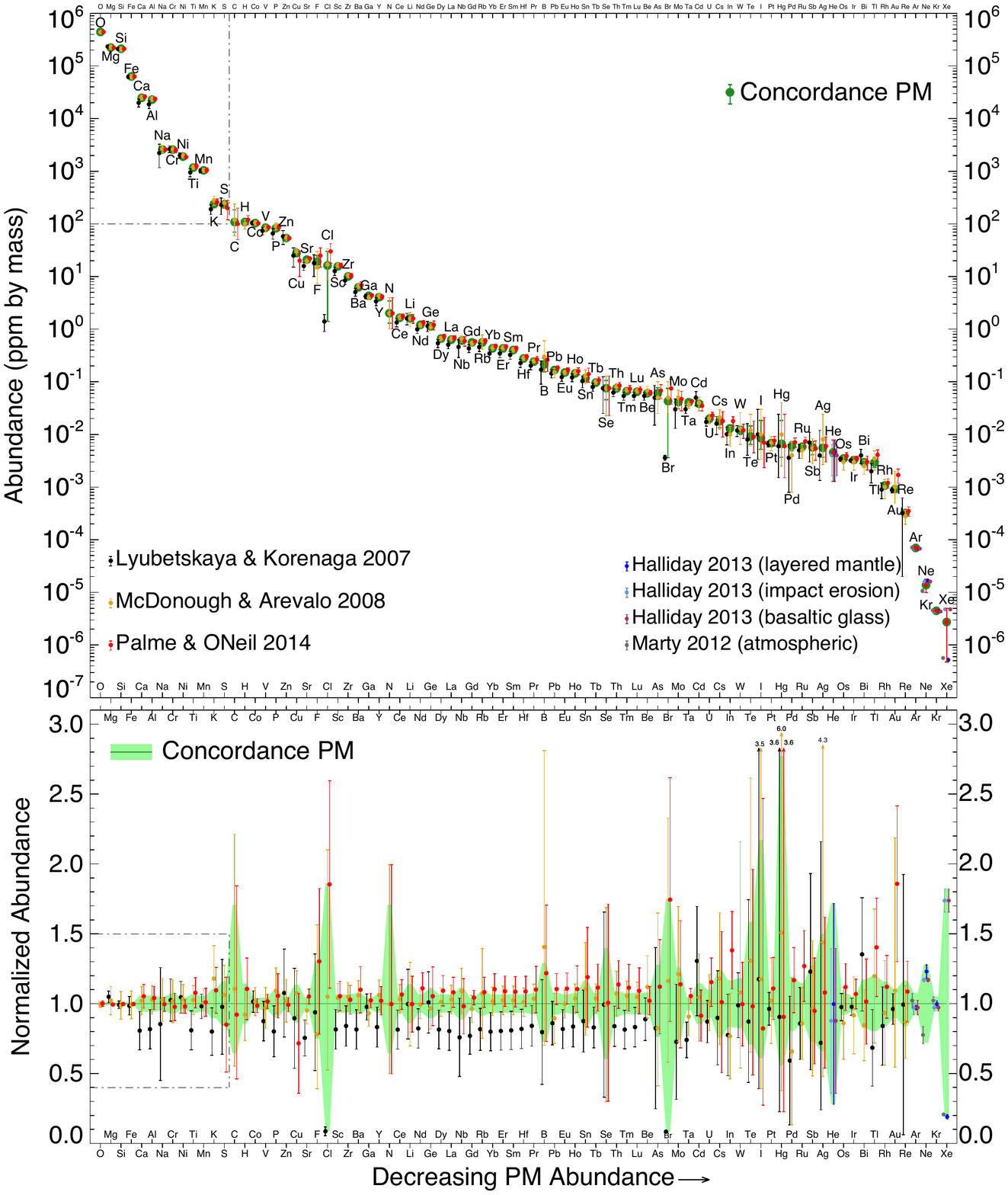} 
		\caption{Recent literature estimates of 83 elemental abundances in the primitive mantle (PM) and our concordance estimates (green) constructed from them. Elements are plotted in order of decreasing PM abundance. The upper panel plots ppm by mass. In the lower panel, literature values have been normalized to our PM concordance estimates. Our PM concordance ppm estimates have been rescaled up by $0.3\%$ to ensure that their sum equals $10^6$. We have not rescaled the literature values (see Appendix C for rescaling details). The light green band in the lower panel indicates our estimate of the uncertainties on the concordance values. $64\%$ of the literature points fall within this band. The string of relatively low values for RLEs in the middle of the plot for the \cite{Lyubetskaya2007} data set is due to its	relatively high Mg abundance. Upper limits that extend beyond the plot range are labeled with their y-values. The dashed boxes on the left in both panels contain the 13 most abundant elements and are zoomed-in on in Fig. \ref{fig:mantle-2}.}
		\label{fig:mantle-1}
	\end{center}
\end{figure}

\begin{figure}[H]
	\begin{center}
		\includegraphics[trim=0.9cm 2.6cm 0.5cm 2.8cm, scale=0.80,angle=0]{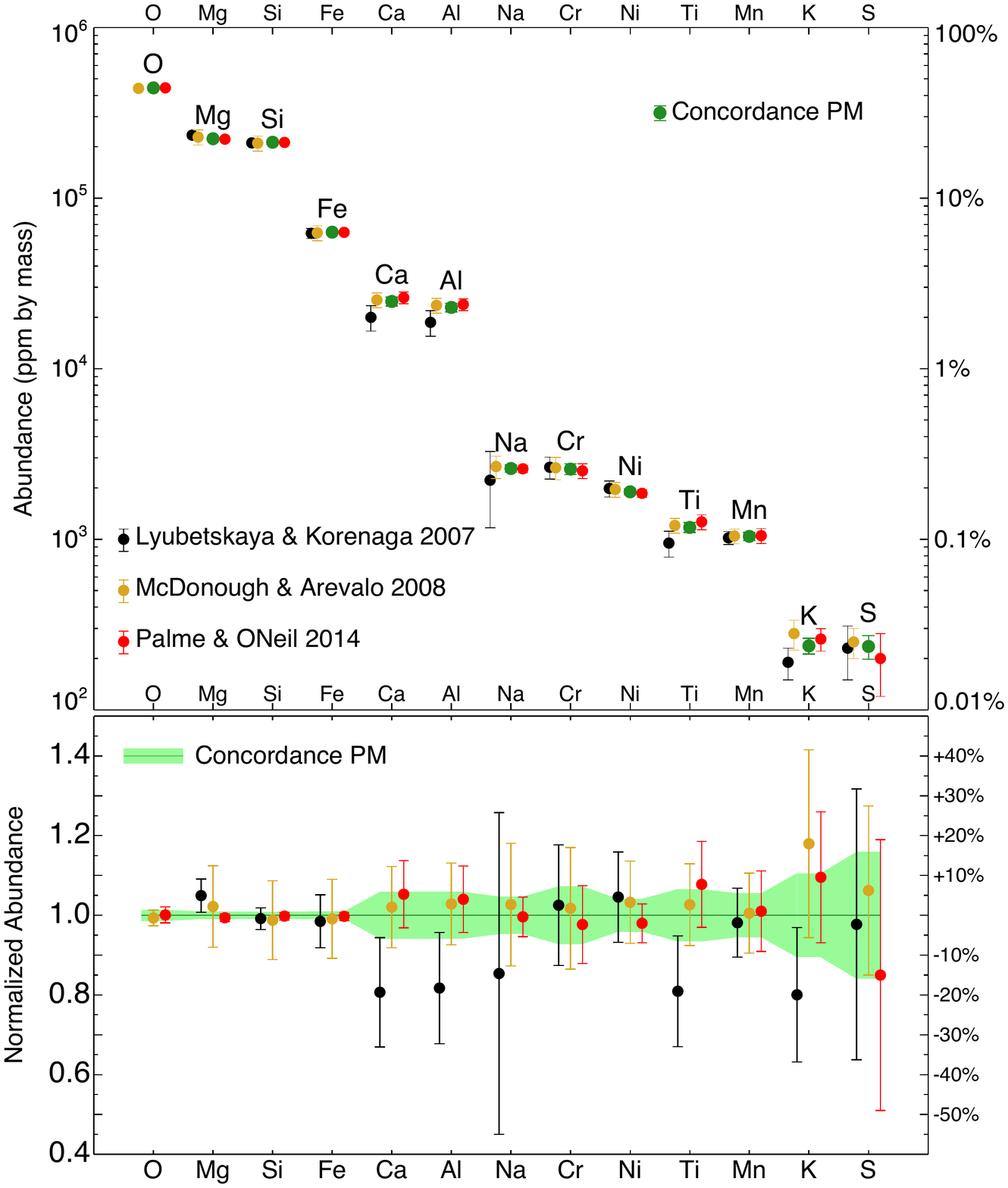} 
		\caption{Zoom-in of the 13 most abundant elements contained in the dashed boxes in both panels of the previous figure. The sum of the abundances of the 4 most abundant elements (O, Mg, Si, Fe) make up $94.19 \pm 0.69 \%$ of the total PM mass. The 6 most abundant elements (O, Mg, Si, Fe, Ca, Al) make up $98.96\pm 0.72 \%$, while the 13 most abundant elements plotted here make up $99.93^{+0.07}_{-0.72} \%$ of the total PM mass. The extra mass in Mg in the \cite{Lyubetskaya2007} data set is compensated for by their low values of Ca and Al . This is a good example of the covariance or non-independence of elemental abundance estimates.}
		\label{fig:mantle-2}
	\end{center}
\end{figure}

The resultant concordance estimates for the PM elemental abundances and their uncertainties, are listed in column 3 of Table \ref{tab:earth}. 
Our PM elemental abundances are increased by $0.3\%$  to ensure that the sum of all the ppm values equals $10^6$ (see \ref{sec:rescale}). 
Figs. \ref{fig:mantle-1} \& \ref{fig:mantle-2} show the comparison of our concordance abundances with the literature values.
The upper panel of Fig. \ref{fig:mantle-1} shows elemental abundances (ppm by mass). 
The literature abundances normalized to our concordance PM abundances are shown in the lower panel. 
For clarity, Fig. \ref{fig:mantle-2} is a zoomed-in version of the 13 most abundant elements in Fig. \ref{fig:mantle-1}.
The abundances of these 13 elements account for $99.93^{+0.07}_{-0.72}$ wt\% of primitive mantle.

By construction, our concordance PM abundances are consistent with the literature abundances. 
There are some outliers. For example, the  \cite{Lyubetskaya2007} abundances of Cl and Br are relatively low because they used the Cl/K ratio of 0.0075$\pm$0.0025 from highly depleted MORB in \cite{Saal2002}. This ratio is $\sim$ 10 times lower than an equivalent Cl/K ratio of $\sim$ 0.07 used in \cite{McDonough2008} \citep[which came from the Cl/Rb $\sim$ 28 and K/Rb $\sim$ 400 of][]{McDonough1995}. The abundances of K estimated in \cite{Lyubetskaya2007} and \cite{McDonough2008} are consistent (190$\pm$76 ppm and 240$\pm$48 ppm, respectively), resulting in the abundance of Cl in \cite{Lyubetskaya2007} $\sim$ 12 times lower than that in \cite{McDonough2008}. Based on a 10\% partial melting MORB and the mass balance, \cite{Palme2014b} estimate a total CI content of 30 ppm for PM, $\sim$ 21 times higher that that ($\sim$ 1.4 ppm) in \cite{Lyubetskaya2007}.   
The \cite{Lyubetskaya2007} abundance of Br comes from the Cl/Br ratio of $\sim$ 400$\pm$50, which is the same ratio used in \cite{Palme2014b} and is higher than the approximate 350 \citep{McDonough2008, McDonough1995}. 
Thus, the \cite{Lyubetskaya2007} Br abundance is also low and inconsistent with that of \cite{McDonough2008} and of \cite{Palme2014b}.
Because of this inconsistency, we use an unweighted mean for Cl and Br (see Appendix A.1). 
As a result, the ratio of our estimated Cl and Br abundances is $\sim$ 376.

\begin{table*}
	{\tiny
		\caption{Concordance estimates of the elemental abundances of primitive mantle, core, and bulk Earth.
			Following convention, depending on the element, we use ppm (by mass) or $wt\%$ or ppb (as indicated in column 2).} 
		\begin{center}
			 \begin{tabular}{ll S[table-format=4.5] l S[table-format=4.5] l S[table-format=4.5]}
				\toprule
				\multirow{2}{*}{Z} & \multirow{2}{*}{Elem} & \multicolumn{2}{l}{\bf Concordance Primitive Mantle} & \multicolumn{2}{l}{\bf Concordance Core } &\multicolumn{1}{l}{\bf Bulk Earth$^d$}\\
				&  & \multicolumn{1}{p{2cm}}{abundance (ppm)$^a$} & {Sources$^b$} & \multicolumn{1}{p{2cm}}{abundance (ppm)$^a$} & \multicolumn{1}{p{2cm}}{Sources$^c$} & {abundance (ppm)$^a$}\\
				\hline  
1	&	H	&	109	$\pm$	15				&	MA08, PO14	&	569	$\pm$	421				&	W06, MA08, ZY12	&	258	$\pm$	137				\\
2	&	He (ppb)	&	4.56	$\pm$	3.29				&	H13	&	{-}						&	{-}	&	3.08	$\pm$	2.22				\\
3	&	Li	&	1.60	$\pm$	0.22				&	LK07, MA08, PO14	&	1.57	$\pm$	0.59				&	KL93	&	1.59	$\pm$	0.24				\\
4	&	Be	&	0.0607	$\pm$	0.0048				&	LK07	&	{-}						&	{-}	&	0.0410	$\pm$	0.0033				\\
5	&	B	&	0.213	$^{+	0.062	}_{-	0.059	}$	&	LK07, MA08, PO14	&	{-}						&	{-}	&	0.144	$^{+	0.042	}_{-	0.040	}$	\\
6	&	C	&	109	$^{+	77	}_{-	39	}$	&	MA08, PO14	&	7921	$\pm$	6820				&	W06, MA08, ZY12, W13, N15, LS16	&	2648	$\pm$	2217				\\
7	&	N	&	2.01	$^{+	1.42	}_{-	0.71	}$	&	MA08, PO14	&	93	$^{+	77	}_{-	90	}$	&	MA08, ZY12	&	31.7	$^{+	24.9	}_{-	29.3	}$	\\
8	&	O (\%)	&	44.3	$\pm$	0.6				&	MA08, PO14	&	2.66	$\pm$	1.82				&	 KL93, A01, M03, J10, H11, R11, ZY12, H13, S13, B14, LS16	&	30.8	$\pm$	0.7				\\
9	&	F	&	19.2	$^{+	5.8	}_{-	4.9	}$	&	LK07, MA08, PO14	&	{-}						&	{-}	&	13.0	$^{+	3.9	}_{-	3.3	}$	\\
10	&	Ne (ppb)	&	0.0137	$\pm$	0.0039				&	M12, H13	&	{-}						&	{-}	&	0.00927	$\pm$	0.00261				\\
11	&	Na	&	2600	$\pm$	123				&	LK07, MA08, PO14	&	1372	$\pm$	519				&	KL93	&	2201	$\pm$	188				\\
12	&	Mg (\%)	&	22.3	$\pm$	0.2				&	LK07, MA08, PO14	&	0.0588	$^{+	0.0294	}_{-	0.0288	}$	&	ZY12	&	15.1	$\pm$	0.2				\\
13	&	Al (\%)	&	2.29	$\pm$	0.13				&	LK07, MA08, PO14	&	{-}						&	{-}	&	1.54	$\pm$	0.09				\\
14	&	Si (\%)	&	21.3	$\pm$	0.2				&	LK07, MA08, PO14	&	4.96	$\pm$	2.34				&	A01, W06, MA08, J10, R11, ZY12, H13, S13, B14, LS16	&	16.0	$\pm$	0.8				\\
15	&	P	&	82.4	$\pm$	8.0				&	LK07, MA08, PO14	&	2774	$^{+	916	}_{-	1236	}$	&	A01, MA08, KL93, ZY12	&	957	$^{+	298	}_{-	402	}$	\\
16	&	S	&	235	$\pm$	38				&	LK07, MA08, PO14	&	18269	$\pm$	4497				&	KL93, A01, W06, MA08, R11, H13, LS16  	&	6096	$\pm$	1463				\\
17	&	Cl	&	16.2	$^{+	13.9	}_{-	14.8	}$	&	LK07, MA08, PO14	&	459	$^{+	278	}_{-	259	}$	&	KL93, MA08	&	160	$^{+	91	}_{-	85	}$	\\
18	&	Ar (ppb)	&	0.0697	$\pm$	0.0054				&	M12, H13	&	{-}						&	{-}	&	0.0471	$\pm$	0.0036				\\
19	&	K	&	237	$\pm$	25				&	LK07, MA08, PO14	&	206	$\pm$	78				&	KL93	&	227	$\pm$	30				\\
20	&	Ca (\%)	&	2.48	$\pm$	0.15				&	LK07, MA08, PO14	&	{-}						&	{-}	&	1.67	$\pm$	0.10				\\
21	&	Sc	&	15.6	$\pm$	1.0				&	LK07, MA08, PO14	&	{-}						&	{-}	&	10.5	$\pm$	0.7				\\
22	&	Ti	&	1174	$\pm$	77				&	LK07, MA08, PO14	&	{-}						&	{-}	&	792	$\pm$	52				\\
23	&	V	&	84.6	$\pm$	3.9				&	LK07, MA08, PO14	&	122	$\pm$	46				&	KL93, MA08, R11	&	96.6	$\pm$	15.2				\\
24	&	Cr	&	2580	$\pm$	187				&	LK07, MA08, PO14	&	6590	$\pm$	1751				&	KL93, A01, MA08, J10, R11	&	3883	$\pm$	583				\\
25	&	Mn	&	1040	$\pm$	57				&	LK07, MA08, PO14	&	3257	$^{+	2563	}_{-	2903	}$	&	KL93, A01, M03, MA08	&	1760	$^{+	834	}_{-	944	}$	\\
26	&	Fe (\%)	&	6.32	$\pm$	0.06				&	LK07, MA08, PO14	&	82.8	$\pm$	2.9				&	KL93, A01,M03, MA08, J10, R11	&	31.2	$\pm$	1.0				\\
27	&	Co	&	103	$\pm$	4				&	LK07, MA08, PO14	&	2373	$\pm$	141				&	KL93, A01, MA08, J10, R11	&	841	$\pm$	46				\\
28	&	Ni (\%)	&	0.190	$\pm$	0.008				&	LK07, MA08, PO14	&	5.06	$\pm$	0.23				&	KL93, A01,M03, MA08, J10, R11	&	1.77	$\pm$	0.08				\\
29	&	Cu	&	27.9	$\pm$	3.8				&	LK07, MA08, PO14	&	154	$\pm$	58				&	MA08, KL93	&	69.0	$\pm$	19.2				\\
30	&	Zn	&	53.9	$\pm$	2.5				&	LK07, MA08, PO14	&	29	$\pm$	11				&	KL93	&	45.9	$\pm$	4.0				\\
31	&	Ga	&	4.30	$\pm$	0.17				&	LK07, MA08, PO14	&	4.90	$\pm$	1.85				&	KL93	&	4.49	$\pm$	0.61				\\
32	&	Ge	&	1.14	$\pm$	0.12				&	LK07, MA08, PO14	&	24.0	$\pm$	9.1				&	KL93, MA08	&	8.56	$\pm$	2.95				\\
33	&	As	&	0.0606	$^{+	0.0167	}_{-	0.0145	}$	&	LK07, MA08, PO14	&	4.89	$\pm$	1.85				&	KL93, MA08	&	1.63	$\pm$	0.60				\\
34	&	Se	&	0.0755	$\pm$	0.0300				&	LK07, MA08, PO14	&	8.63	$\pm$	3.27				&	KL93, MA08	&	2.86	$\pm$	1.06				\\
35	&	Br	&	0.0430	$^{+	0.0322	}_{-	0.0394	}$	&	LK07, MA08, PO14	&	0.514	$\pm$	0.195				&	KL93, MA08	&	0.196	$^{+	0.067	}_{-	0.069	}$	\\
36	&	Kr (ppb)	&	0.00446	$\pm$	0.00022				&	M12, H13	&	{-}						&	{-}	&	0.00301	$\pm$	0.00015				\\
37	&	Rb	&	0.559	$\pm$	0.048				&	LK07, MA08, PO14	&	1.08	$\pm$	0.41				&	KL93	&	0.728	$\pm$	0.136				\\
38	&	Sr	&	20.9	$\pm$	0.9				&	LK07, MA08, PO14	&	{-}						&	{-}	&	14.1	$\pm$	0.6				\\
39	&	Y	&	4.05	$\pm$	0.27				&	LK07, MA08, PO14	&	{-}						&	{-}	&	2.73	$\pm$	0.18				\\
40	&	Zr	&	10.02	$\pm$	0.66				&	LK07, MA08, PO14	&	{-}						&	{-}	&	6.76	$\pm$	0.44				\\
41	&	Nb	&	0.606	$\pm$	0.070				&	LK07, MA08, PO14	&	0.423	$\pm$	0.160				&	R11	&	0.547	$\pm$	0.070				\\
42	&	Mo	&	0.0413	$\pm$	0.0107				&	LK07, MA08, PO14	&	4.97	$\pm$	1.88				&	KL93, MA08	&	1.64	$\pm$	0.61				\\
44	&	Ru (ppb)	&	5.83	$\pm$	0.86				&	LK07, MA08, PO14	&	3939	$\pm$	1491				&	KL93, MA08	&	1284	$\pm$	485				\\
45	&	Rh (ppb)	&	1.07	$\pm$	0.17				&	LK07, MA08, PO14	&	684	$\pm$	259				&	KL93, MA08	&	223	$\pm$	84				\\
46	&	Pd (ppb)	&	6.08	$\pm$	1.18				&	LK07, MA08, PO14	&	2719	$\pm$	1029				&	KL93, MA08	&	888	$\pm$	335				\\
47	&	Ag (ppb)	&	5.56	$^{+	2.79	}_{-	1.89	}$	&	LK07, MA08, PO14	&	239	$\pm$	90				&	KL93, MA08	&	81.3	$\pm$	29.4				\\
48	&	Cd	&	0.0383	$\pm$	0.0056				&	LK07, MA08, PO14	&	0.167	$\pm$	0.063				&	KL93, MA08	&	0.0800	$\pm$	0.0208				\\
49	&	In	&	0.0130	$\pm$	0.0022				&	LK07, MA08, PO14	&	{-}						&	{-}	&	0.00879	$\pm$	0.00148				\\
50	&	Sn	&	0.118	$\pm$	0.019				&	LK07, MA08, PO14	&	0.490	$\pm$	0.185				&	KL93, MA08	&	0.239	$\pm$	0.062				\\
51	&	Sb (ppb)	&	5.69	$\pm$	1.57				&	LK07, MA08, PO14	&	132	$\pm$	50				&	KL93, MA08	&	46.8	$\pm$	16.3				\\
52	&	Te (ppb)	&	9.18	$^{+	5.37	}_{-	2.68	}$	&	LK07, MA08, PO14	&	1009	$\pm$	382				&	KL93, MA08	&	334	$\pm$	124				\\
53	&	I (ppb)	&	8.51	$^{+	9.98	}_{-	3.33	}$	&	LK07, MA08, PO14	&	92.1	$\pm$	34.9				&	KL93, MA08	&	35.7	$^{+	13.2	}_{-	11.6	}$	\\
54	&	Xe (ppb)	&	0.00274	$\pm$	0.00226				&	M12, H13	&	{-}						&	{-}	&	0.00185	$\pm$	0.00152				\\
55	&	Cs	&	0.0178	$\pm$	0.0043				&	LK07, MA08, PO14	&	0.103	$\pm$	0.039				&	KL93, MA08	&	0.0455	$\pm$	0.0130				\\
56	&	Ba	&	6.23	$\pm$	0.47				&	LK07, MA08, PO14	&	{-}						&	{-}	&	4.21	$\pm$	0.32				\\
57	&	La	&	0.632	$\pm$	0.042				&	LK07, MA08, PO14	&	{-}						&	{-}	&	0.427	$\pm$	0.028				\\
58	&	Ce	&	1.65	$\pm$	0.11				&	LK07, MA08, PO14	&	{-}						&	{-}	&	1.11	$\pm$	0.07				\\
59	&	Pr	&	0.241	$\pm$	0.018				&	LK07, MA08, PO14	&	{-}						&	{-}	&	0.163	$\pm$	0.012				\\
60	&	Nd	&	1.21	$\pm$	0.08				&	LK07, MA08, PO14	&	{-}						&	{-}	&	0.815	$\pm$	0.054				\\
62	&	Sm	&	0.401	$\pm$	0.026				&	LK07, MA08, PO14	&	{-}						&	{-}	&	0.270	$\pm$	0.018				\\
63	&	Eu	&	0.150	$\pm$	0.010				&	LK07, MA08, PO14	&	{-}						&	{-}	&	0.101	$\pm$	0.007				\\
64	&	Gd	&	0.561	$\pm$	0.024				&	LK07, MA08, PO14	&	{-}						&	{-}	&	0.379	$\pm$	0.017				\\
65	&	Tb	&	0.0964	$\pm$	0.0073				&	LK07, MA08, PO14	&	{-}						&	{-}	&	0.0651	$\pm$	0.0049				\\
66	&	Dy	&	0.662	$\pm$	0.043				&	LK07, MA08, PO14	&	{-}						&	{-}	&	0.447	$\pm$	0.029				\\
67	&	Ho	&	0.145	$\pm$	0.011				&	LK07, MA08, PO14	&	{-}						&	{-}	&	0.0976	$\pm$	0.0074				\\
68	&	Er	&	0.430	$\pm$	0.028				&	LK07, MA08, PO14	&	{-}						&	{-}	&	0.290	$\pm$	0.019				\\
69	&	Tm	&	0.0663	$\pm$	0.0050				&	LK07, MA08, PO14	&	{-}						&	{-}	&	0.0447	$\pm$	0.0034				\\
70	&	Yb	&	0.433	$\pm$	0.028				&	LK07, MA08, PO14	&	{-}						&	{-}	&	0.292	$\pm$	0.019				\\
71	&	Lu	&	0.0649	$\pm$	0.0049				&	LK07, MA08, PO14	&	{-}						&	{-}	&	0.0438	$\pm$	0.0033				\\
72	&	Hf	&	0.277	$\pm$	0.018				&	LK07, MA08, PO14	&	{-}						&	{-}	&	0.187	$\pm$	0.012				\\
73	&	Ta	&	0.0408	$\pm$	0.0019				&	LK07, MA08, PO14	&	0.00799	$^{+	0.00407	}_{-	0.00399	}$	&	R11	&	0.0301	$\pm$	0.0018				\\
74	&	W	&	0.0120	$^{+	0.0022	}_{-	0.0021	}$	&	LK07, MA08, PO14	&	0.502	$\pm$	0.190				&	KL93, MA08, R11	&	0.171	$\pm$	0.062				\\
75	&	Re (ppb)	&	0.322	$\pm$	0.053				&	LK07, MA08, PO14	&	214	$\pm$	81				&	KL93, MA08	&	69.8	$\pm$	26.4				\\
76	&	Os (ppb)	&	3.49	$\pm$	0.28				&	LK07, MA08, PO14	&	2719	$\pm$	1029				&	KL93, MA08	&	886	$\pm$	335				\\
77	&	Ir (ppb)	&	3.27	$\pm$	0.17				&	LK07, MA08, PO14	&	2611	$\pm$	989				&	KL93, MA08	&	851	$\pm$	321				\\
78	&	Pt (ppb)	&	6.86	$\pm$	0.67				&	LK07, MA08, PO14	&	5448	$\pm$	2063				&	KL93, MA08	&	1775	$\pm$	671				\\
79	&	Au (ppb)	&	0.915	$\pm$	0.098				&	LK07, MA08, PO14	&	481	$\pm$	182				&	KL93, MA08	&	157	$\pm$	59				\\
80	&	Hg (ppb)	&	6.63	$^{+	11.75	}_{-	2.94	}$	&	LK07, MA08, PO14	&	32.3	$^{+	17.7	}_{-	16.3	}$	&	KL93, MA08	&	15.0	$^{+	9.8	}_{-	5.7	}$	\\
81	&	Tl (ppb)	&	2.92	$\pm$	0.58				&	LK07, MA08, PO14	&	29.4	$\pm$	11.1				&	MA08	&	11.5	$\pm$	3.6				\\
82	&	Pb	&	0.167	$\pm$	0.014				&	LK07, MA08, PO14	&	1.715	$^{+	1.585	}_{-	1.515	}$	&	KL93, MA08	&	0.670	$^{+	0.515	}_{-	0.492	}$	\\
83	&	Bi (ppb)	&	2.96	$\pm$	0.52				&	LK07, MA08, PO14	&	19.1	$^{+	10.9	}_{-	10.1	}$	&	KL93, MA08	&	8.21	$^{+	3.56	}_{-	3.30	}$	\\
90	&	Th	&	0.0746	$\pm$	0.0068				&	LK07, MA08, PO14	&	{-}						&	{-}	&	0.0504	$\pm$	0.0046				\\
92	&	U	&	0.0198	$\pm$	0.0020				&	LK07, MA08, PO14	&	{-}						&	{-}	&	0.0134	$\pm$	0.0013				\\
\hline																												
	&	Total	&	1.00	$\times10^6$					&		&	1.00	$\times10^6$					&		&	1.00	$\times10^6$					\\
				\bottomrule
				\multicolumn{7}{p{14cm}}{\tiny $^a$ ppm (by mass) unless otherwise indicated in column 2. Values have been rescaled $\sum_{\mathrm{ppm}} = 10^6$, see Table \ref{tab:S2}.}\\
				\multicolumn{7}{p{14cm}}{\tiny $^b$ LK07: \cite{Lyubetskaya2007} columns 4 and 5 of their Table 3; 
					MA08: \cite{McDonough2008} columns 2 and 6 of their Table 1; 
					PO14: \cite{Palme2014b} columns 5 and 6 of their Table 4; 
					H13: \cite{Halliday2013} columns 7, 10 and 13 of their Table 2; 
					M12: \cite{Marty2012} last column of his Table 1. 
					See Sect. \ref{sec:pm-2} for the details of our concordance PM abundances.}\\
				\multicolumn{7}{p{14cm}}{\tiny $^c$ for the key to the literature acronyms for core abundances, see footnote a of Table 2.}\\
				\multicolumn{7}{p{14cm}}{\tiny $^d$ Weighted sum (Eqs. \ref{eq:5} \& \ref{eq:6}) of concordance estimates of the abundances of primitive mantle and core.}\\
				
			\end{tabular}
		\end{center}
		\label{tab:earth}
	}  
\end{table*}
\clearpage
\section{Composition of the Core}
\label{sec:core}
\subsection{Data sources}
\label{sec:core-1}

The density deficit of the Earth's core (compared to a pure Fe-Ni composition) suggests that the core contains a significant amount of one or more light elements. 
The liquid outer core is thought to have a density deficit of  $3-12$ wt\%  \citep[e.g.,][]{Stevenson1981, Anderson1994, Anderson2002, McDonough2003}, while the solid inner core is 3-6 wt\% less dense than predicted \citep[e.g.,][]{Anderson1994, Hemley2001}. 
The candidate light elements are still controversial and have been modeled and estimated in various ways and plausibly include Si, O, S and C \citep[see reviews by][]{Hirose2013, Litasov2016}. \\[6pt]
We have compiled and combined a wide variety of recent work to construct concordance core abundances.
These are listed in Tables 1 and 2 and include various constraints from various core compositional models, terrestrial fractionation curves and mass balance \citep[e.g.][]{Kargel1993, Allegre1995, McDonough2003, Wood2006, McDonough2008}, metal-silicate equilibrium \citep[e.g.][]{Rubie2011}, the chemistry of core formation \citep[e.g.][]{Javoy2010}, high-pressure and high-temperature experiments \citep{Siebert2013}, experiments based on sound velocity and/or density jumps \citep{Huang2011, Nakajima2015}, and numerical simulations \citep[e.g.][]{Zhang2012, Badro2014}.

\begin{table*}
	{\tiny
		\centering
		\caption{Concentrations (wt\%) of the 13 most abundant elements in the core}
		\begin{tabular}{l ll lllllllllll  }
			\toprule
			\multicolumn{14}{c}{} \\
			Sources$^a$ & Fe & Ni & Si & O & S & C & Cr & Mn & P & Co & Na & Mg & H \\
			\hline  
			KL93	&	85.55	&	4.88	&	-	&	5.18	&	2.69	&	-	&	0.45	&	0.41 &	0.347		&	0.218	&	0.14	&	-	&	-	\\
			A01/95	&	79.4$\pm$2.0	&	4.87$\pm$0.30	&	7.35	&	5.0$\pm$0.5	&	1.21$\pm$0.20	&	-	&	0.78	&	0.582 &	0.369		&	0.253	&	-	&	-	&	-	\\
			M03	&	88.3	&	5.4	&	nil	&	3	&	-	&	-	&	-	&	0.03 &	-		&	-	&	-	&	-	&	-	\\
			W06	&	-	&	-	&	4.5$\pm$0.5	&	-	&	1.9	&	0.2	&	-	&	-	&	-	&	-	&	-	&	-	&	0.1	\\
			MA08	&	85	&	5.2	&	6.4	&	nil	&	1.9	&	0.2	&	0.9	&	0.005 &	0.32		&	0.25	&	-	&	-	&	0.06	\\
			J10	&	85.5$\pm$1.1	&	5.35$\pm$0.81	&	6.64$\pm$0.51	&	1.99$\pm$0.46	&	-	&	-	&	0.55$\pm$0.05	&	-	&	-	&	0.25$\pm$0.03	&	-	&	-	&	-	\\
			H11	&	-	&	-	&	-	&	0.5	&	-	&	-	&	-	&	-	&	-	&	-	&	-	&	-	&	-	\\
			R11	&	83.45$\pm$0.35	&	5.3$\pm$0.1	&	8.4$\pm$0.2	&	0.655$\pm$0.185	&	2	&	-	&	0.68$\pm$0.12	&	-	&	-	&	0.24$\pm$0.01	&	-	&	-	&	-	\\
			ZY12	&	-	&	-	&	2.0$\pm$0.2	&	1.0$\pm$0.1	&	-	&	0.4$\pm$0.3	&	-	&	- &	0.096$\pm$0.060		&	-	&	-	&	0.06$\pm$0.03	&	0.014$\pm$0.008	\\
			H13	&	-	&	-	&	6	&	3	&	1.5$\pm$0.5	&	-	&	-	&	-	&	-	&	-	&	-	&	-	&	-	\\
			W13	&	-	&	-	&	-	&	-	&	-	&	1	&	-	&	-	&	-	&	-	&	-	&	-	&	-	\\
			S13	&	-	&	-	&	1.85$\pm$0.35	&	5.0$\pm$0.5	&	-	&	-	&	-	&	-	&	-	&	-	&	-	&	-	&	-	\\
			B14	&	-	&	-	&	1.94$\pm$0.21	&	3.8$\pm$0.7	&	nil	&	nil	&	-	&	-	&	-	&	-	&	-	&	-	&	-	\\
			N15	&	-	&	-	&	-	&	-	&	-	&	1.05$\pm$0.15	&	-	&	-	&	-	&	-	&	-	&	-	&	-	\\
			LS16	&	-	&	-	&	5.5$\pm$0.5	&	0.75$\pm$0.25	&	1.85$\pm$0.05	&	2	& -		&	-	&	-	&	-	&	-	&	-	&	-	\\
			{\bf Mean$^b$}	&	{\bf 82.8$\pm$2.9}	&	{\bf 5.06$\pm$0.23}	&	{\bf 4.96$\pm$2.34}	&	{\bf 2.66$\pm$1.82}	&	{\bf 1.83$\pm$0.45}	&	{\bf 0.79$\pm$0.68}	&	{\bf 0.66$\pm$0.18}	&	{\bf0.326$^{+0.256}_{-0.290}$}	&	{\bf 0.277$^{+0.092}_{-0.124}$}	&	{\bf 0.237$\pm$0.014} &	{\bf 0.14$\pm$0.05$^c$ }	&	{\bf 0.06$\pm$0.03 }	&	{\bf 0.057$\pm$0.042}	\\
			\bottomrule
			\\
			\multicolumn{14}{p{16.5cm}}{\tiny $^a$ K93: \cite{Kargel1993}: column 6 of Table II}\\			
			\multicolumn{14}{p{16.5cm}}{\tiny A95/A01: \cite{Allegre1995} Table 2 updated with \cite{Allegre2001} for S and O to 1.21 wt\%  and 5 wt\% respectively.}\\
			\multicolumn{14}{p{16.5cm}}{\tiny M03: \cite{McDonough2003}: contents of Fe, O, Si, and Ni in O-bearing calculations listed in Table 7.}\\
			\multicolumn{14}{p{16.5cm}}{\tiny W06: \cite{Wood2006}: last paragraph of Conclusions,  4-5 wt\% Si, 1.9 wt\% S, 0.1 wt\% H and 0.2 wt\% C}\\
			\multicolumn{14}{p{16.5cm}}{\tiny MA08: \cite{McDonough2008} columns 3 and 7 of Table 1}\\
			\multicolumn{14}{p{16.5cm}}{\tiny J10: \cite{Javoy2010} column 4 of Table 6}\\ 
			\multicolumn{14}{p{16.5cm}}{\tiny H11: \cite{Huang2011}: the optimal value of 0.5 wt\% oxygen in the liquid outer core}\\ 
			\multicolumn{14}{p{16.5cm}}{\tiny R11: \cite{Rubie2011}: average of the upper and lower limits of their three heterogeneous models listed in the last three columns of Table 2}\\ 
			\multicolumn{14}{p{16.5cm}}{\tiny ZY12: \cite{Zhang2012}: abundances of O, Mg, and Si are from the M52 simulation results for the core listed in Table 2; abundances of H, C, P, and N are the average of the upper and lower limits of their models A and B}\\ 
			\multicolumn{14}{p{16.5cm}}{\tiny H13: \cite{Hirose2013}: \textquoteleft preferred value' listed in Table 1}\\ 
			\multicolumn{14}{p{16.5cm}}{\tiny W13: \cite{Wood2013}: Conclusions, of $\sim$ 1 wt\% of carbon in the core}\\ 
			\multicolumn{14}{p{16.5cm}}{\tiny S13: \cite{Siebert2013}: conclusion of an oxygen-rich core with 4.5 to 5.5 wt\% O and 1.5 to 2.2 wt\% Si}\\ 
			\multicolumn{14}{p{16.5cm}}{\tiny B14: \cite{Badro2014}: the best numerical fit indicated in Fig.2}\\ 
			\multicolumn{14}{p{16.5cm}}{\tiny N15: \cite{Nakajima2015}: conclusion of 0.9-1.2 wt\% carbon in the core to match sound velocity}\\ 
			\multicolumn{14}{p{16.5cm}}{\tiny LS16 \cite{Litasov2016}: a review of the composition of Earth's core with their best estimates for Si, O, S, and C.}\\
			\multicolumn{14}{p{16.5cm}}{\tiny $^b$ Unweighted mean  of literature values (Eq. \ref{eq:3}), excluding nil estimates; the reported uncertainty is the standard deviation (Eq. \ref{eq:4}).}\\
			\multicolumn{14}{p{16.5cm}}{\tiny $^c$ The uncertainty of 0.05 is assigned by the average uncertainty of 40\% of the 10 most abundant elements in the core (Appendix A).  
			} \\
		\end{tabular}
		\label{tab:core-1}
	} 
\end{table*}
\subsection{Concordance Core Estimate}
\label{sec:core-2}
Since many core abundances reported in the literature are model-dependent and are given without uncertainties, our concordance 
abundances (Tables 1 and 2) are unweighted averages (Eq. \ref{eq:3}) and exclude abundances that have been set to zero. Based on mass balance between the core and the silicate Earth, \cite{McDonough2003} proposed
both Si-bearing and O-bearing core models while \cite{McDonough2008} presented a Si-bearing model only.
The trace elements in the \cite{McDonough2003} O-bearing model and \cite{McDonough2008} Si-bearing model are highly correlated.
Therefore we only count them once in our calculations.
The details of how the literature values were combined into our concordance values with uncertainties are described in \ref{sec:method2}.

\begin{figure}[H]
	\begin{center}
		\includegraphics[trim=0.9cm 2.5cm 1.0cm 2.8cm, scale=0.8,angle=0]{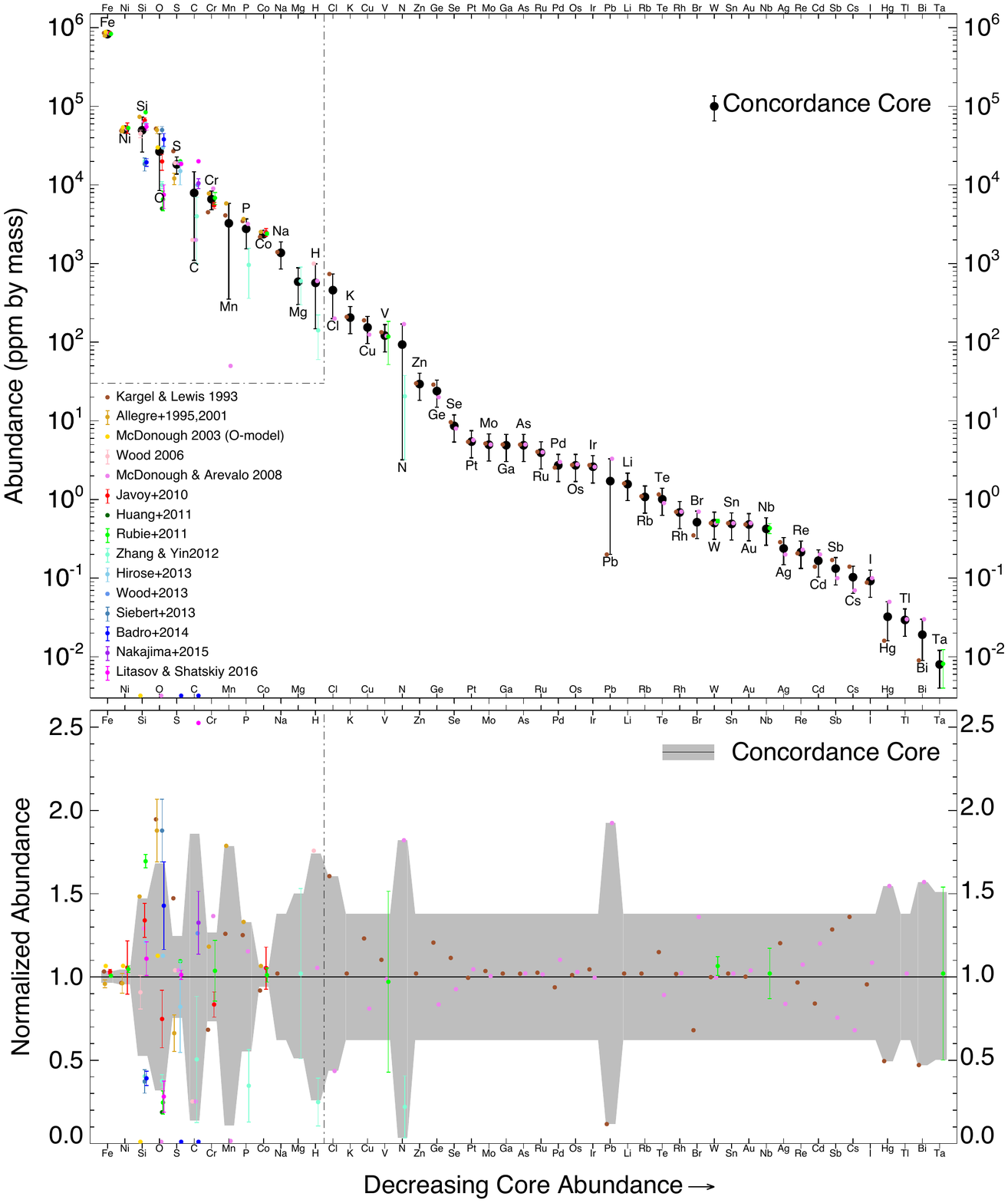} 
		\caption{Recent literature estimates of 49 elemental abundances in the Earth's core and our concordance estimates constructed from them. Elements are plotted in order of decreasing core abundance. The upper panel plots ppm by mass. In the lower panel, literature values have been normalized to our core concordance estimates. The grey band in the lower panel indicates our estimate of the uncertainties on the concordance values. $79\%$ of the literature points fall within this band. The dashed boxes on the left in both panels contain the 13 most abundant elements and are zoomed-in on in Fig. \ref{fig:core-2}. Our core concordance ppm estimates have been rescaled down by $2.0\%$ to constrain their sum to equal $10^6$. We have not rescaled the literature values (see Appendix C, Table C.4).}
		\label{fig:core-1}
	\end{center}
\end{figure}

\begin{figure}[H]
	\begin{center}
		\includegraphics[trim=0.9cm 2.8cm 1.0cm 2.8cm, scale=0.80,angle=0]{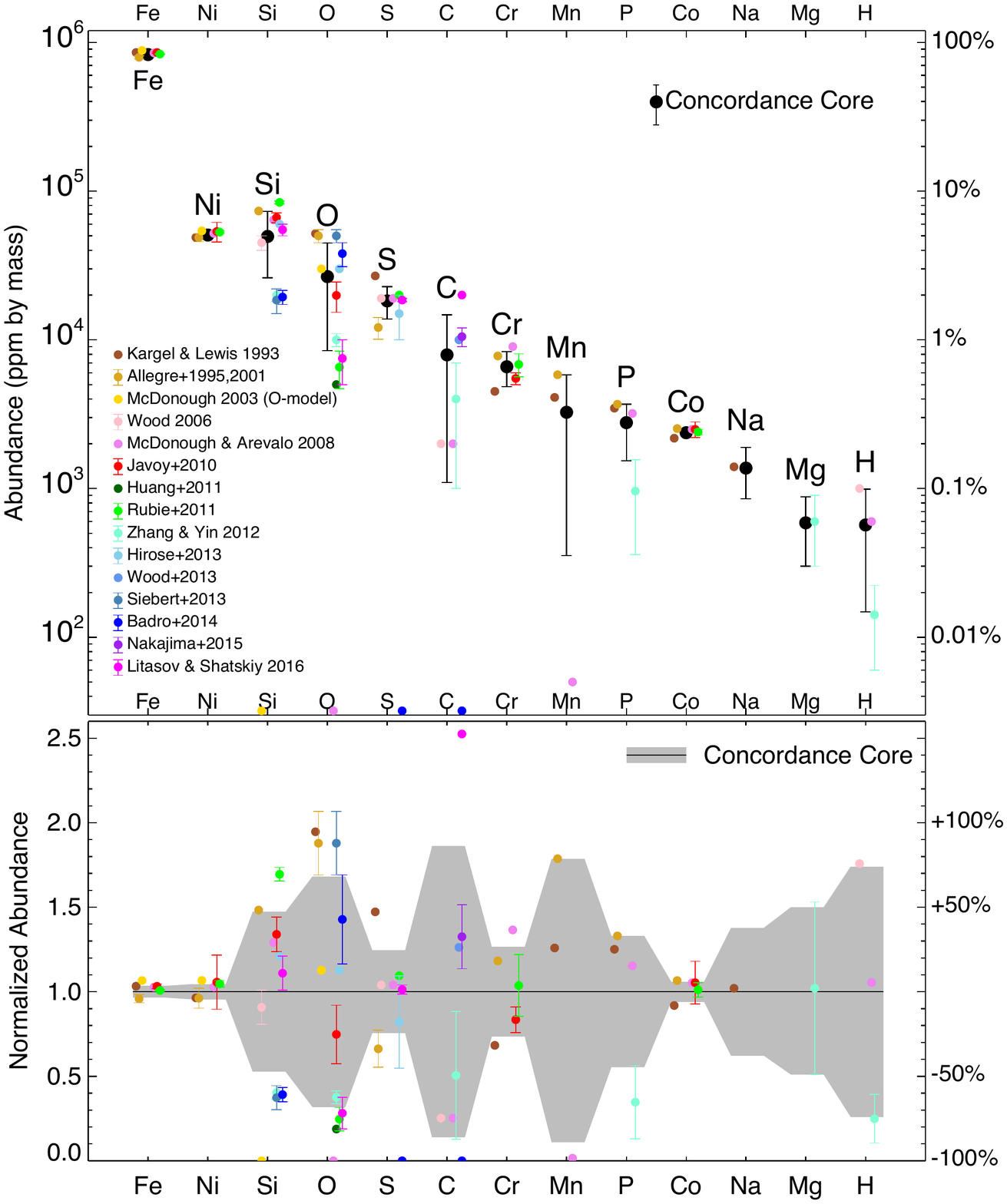} 
		\caption{Zoom-in of the 13 most abundant elements contained in the dashed boxes in both panels of the previous figure. The sum of the Fe and Ni abundances make up $87.90 \pm 2.92 \%$ of the total core mass.The 5 most abundant elements (Fe, Ni, Si, O, S) make up $97.34^{+2.66}_{-4.18} \%$, while the 13 most abundant elements plotted here make up $99.89^{+0.11}_{-4.25} \%$ of the total core mass. See Table 1 for details. The uncertainties on Si and O are not independent since to account for the density deficit in the core, high values of Si probably coincide with low values of O, and vice versa.}
		\label{fig:core-2}
	\end{center}
\end{figure}

Fig. \ref{fig:core-1} shows our concordance abundances for 49 elements in the core, compared with the literature values from which they were constructed. 
The sum of our concordance abundances of the 49 elements is scaled to $10^6$ and listed in column 5 of Table \ref{tab:earth}. 
This is the most complete compilation for the core's composition to date.
Table \ref{tab:core-1} lists the mean concordance concentrations of the 13 most abundant elements in the core along with the literature values
from which they were constructed.  Fig. \ref{fig:core-2} is a zoom-in of the 13 most abundant elements and shows this comparison in more detail.
Fe-Ni alloy accounts for 87.90$\pm$2.92 wt\% of the total mass of the core.  The most abundant light element in the core is Si, followed by O, S, and C. The other less abundant elements include Cr, P, Mn, Co, Na, Mg and H.
%
\section{Composition of the Bulk Earth}
\label{sec:earth}

\subsection{The Core Mass Fraction}
\label{sec:earth-1}

\begin{table*}
	{\scriptsize
		\centering
		\caption{Core mass fraction in the Earth}
		\begin{center}
			\begin{tabular}{l  l  p{9cm}}
				\toprule
				\multicolumn{1}{c}{} \\
				Reference            & Core mass fraction (wt\%) & Comments \\[3pt]
				\hline  
				\cite{Birch1964}     & 32.4 and 32.7 & solutions from two sets of density-velocity relations \\[3pt]
				\cite{Anderson1967}  & 32.5      &  see also \cite{Anderson1989}, using density profiles from \cite{Dziewonski1981} Preliminary Reference Earth Model (PREM model)  based on seismological constraints. \\[3pt]
				\cite{Yoder1995}     & 32.3      &  derived from estimates of the masses of inner core, outer core, and the Earth: $96.75 \times 10^{21}$ kg, $1835 \times 10^{21}$ kg, and $5973.6 \times 10^{21}$ kg respectively; The GEM-T2 Gravitation Model \citep{Marsh1990} and the PREM model \citep{Dziewonski1981} are cited.\\[3pt]
				\cite{Allegre1995,Allegre2001}   & 32.5      &  derived from assumed masses of primitive mantle and core: $4090 \times 10^{21}$ kg and $1967\times 10^{21}$ kg, respectively. Thus a total Earth mass of $6057 \times 10^{21}$ is assumed; \cite{Anders1977}, \cite{Morgan1980} and \cite{Wanke1988} are cited.\\[3pt]
				\cite{McDonough2003} & 32.3      &  cites \cite{Yoder1995}\\[3pt]
				\cite{Javoy2010}     & 32.4      &  no citations given\\[3pt]
				\cite{Hirose2013}    & 33        &  no citations given\\[3pt]
				\cite{Zeng2015}      & 32.5      &  no citations given\\[3pt]
				\bf This work        & {\bf 32.5 $\pm$ 0.3} &  seismological constraints based on the \cite{Dziewonski1981} PREM model, the \cite{Kennett1995} ak135 model, and the radii of inner core and outer core: $3480 \pm 1$ km and $1218 \pm 3$ km, respectively, from \cite{Souriau2015}, along with the constraint of total mass of the Earth ($5972.2 \pm  0.6 \times 10^{21}$ kg, \textit{the Astronomical Almanac Online, USNO-UKHO}.). See details in Section \ref{sec:earth-1}.\\[3pt]
				\bottomrule
				\\
			\end{tabular}
		\end{center}
		\label{tab:earth-1}
	}
\end{table*}
The elemental abundances of bulk Earth are the weighted average of the elemental abundances of primitive mantle and core. 
The accuracy of bulk Earth abundances therefore depends on the accuracy of the core mass fraction.
However, literature values of the core mass fraction (Table \ref{tab:earth-1}) vary, and none has a reported  uncertainty.

Our calculation of the core mass fraction is based on the two standard Earth radial density profile models: PREM \citep{Dziewonski1981} and ak135 \citep{Kennett1995}. 
We take 3480$\pm$1 km for the core-mantle boundary (CMB) and 1218$\pm$3 km for the inner core boundary (ICB) \citep{Souriau2015}. 
We take 5972.2$\pm$0.6 $\times10^{21}$ kg as the total mass of the Earth\footnote{ \textquotedblleft 2016 Selected Astronomical Constants" in \textit{The Astronomical Almanac Online}, USNO-UKHO.}. 
This recent update of the Earth's mass is due to more precise (and accurate) estimates of Newton's constant.
We renormalize the radial density profiles of core and mantle to this new lower Earth mass, which results in an
overall density lower by 0.00158 g/cm$^3$.
We assume a typical uncertainty of 1\% in the density profile models (Brian Kennett, personal communication). 
We include the correlation between core and mantle densities, the uncertainties of the ICB and the CMB, and the fact that an increase of radius can be compensated by a lower density.
Thus, we obtain estimates for the mass fraction of the inner core ($f_{ic}$): 
$f_{ic} = 1.630\pm0.004$ wt\% and of the outer core ($f_{oc}$): $f_{oc} = 30.840\pm0.296$ wt\%. 
The mass fraction of the total core is $f_{ic} + f_{oc} = f_{core}= 32.5 \pm0.3$ wt\%. 
Correspondingly, the weighting factors for estimating the bulk Earth composition are $f_{core}$ and $1-f_{core}$ for the core and PM respectively (Eq. \ref{eq:5}).
\subsection{Concordance Bulk Earth Estimate}
\label{sec:earth-2}
\begin{figure}[H]
	\begin{center}
		\begin{subfigure}[a]{\textwidth}
			\centering
		\includegraphics[trim=1.5cm 1.5cm 1.5cm 2.0cm, scale=0.51,angle=90]{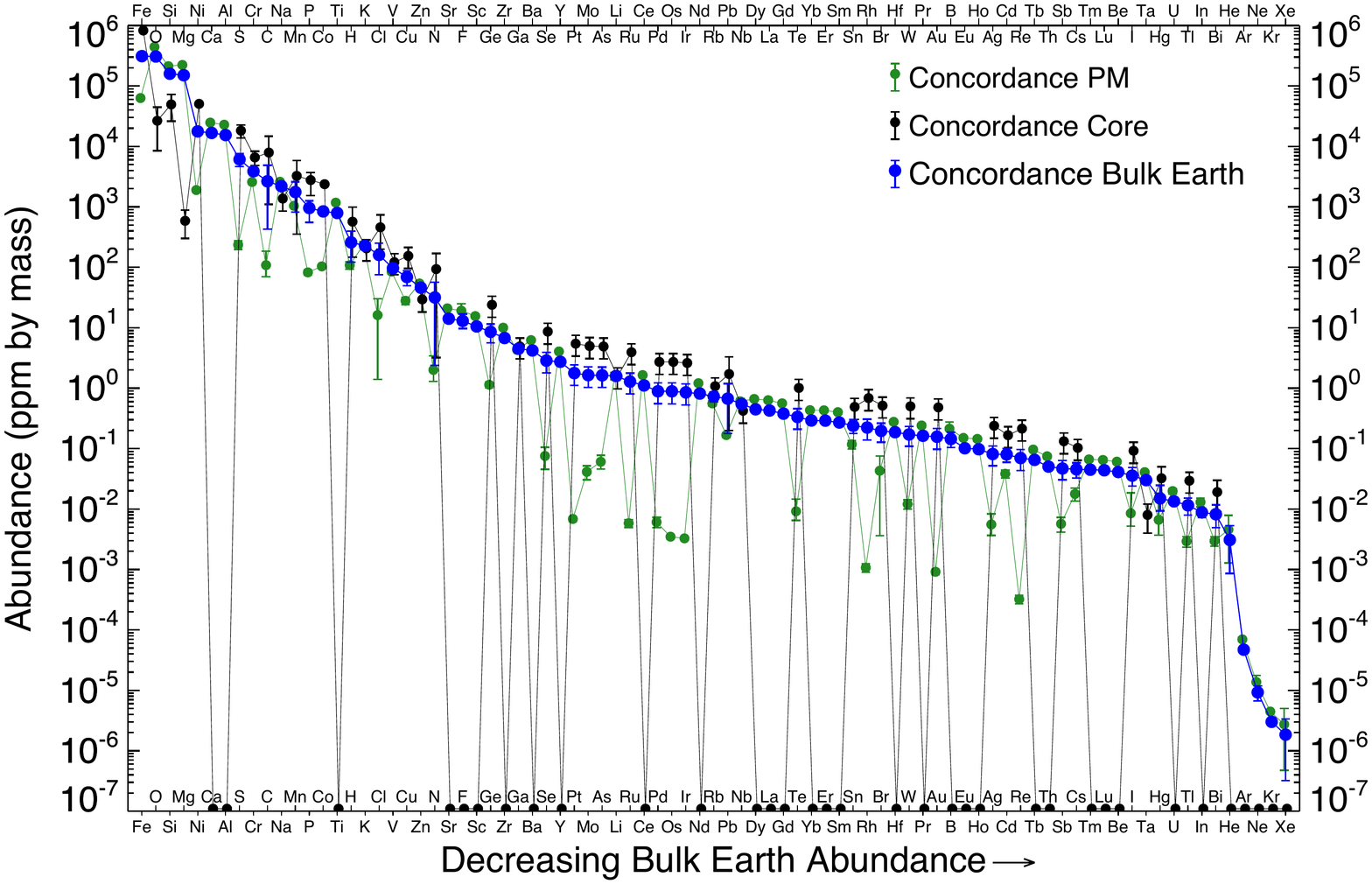} 
		\caption{}
		\label{fig:earth-1}
		\end{subfigure}
        \begin{subfigure}[a]{\textwidth}
        	\centering
		\includegraphics[trim=1.5cm 1.2cm 1.2cm 1.8cm, scale=0.51,angle=90]{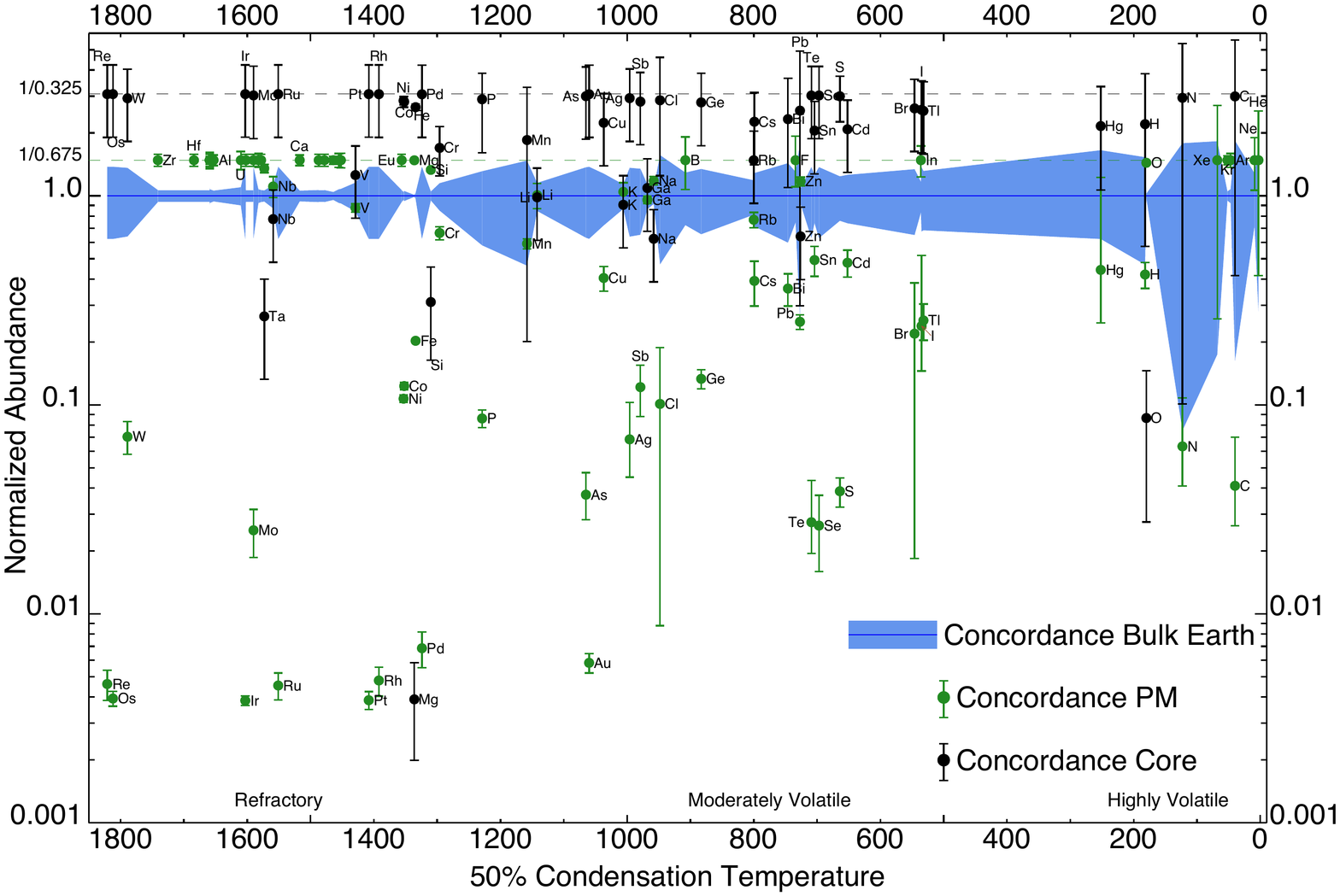} 
		\caption{}
		\label{fig:earthTc}
		\end{subfigure}
\caption{(a) The weighted sums of the concordance PM and the concordance core produce our concordance estimates of elemental abundances of the bulk Earth. The core abundances are weighted by our estimate of the core mass fraction: $32.5 \pm 0.3$ wt\%. The PM abundances are weighted by $67.5 \pm 0.3$ wt\% $ (= 100 - 32.5)$.  Elements with nil or extremely low abundance in the core are plotted on the x-axis. (b) Concordance PM and core abundances normalized to the concordance bulk Earth abundances and plotted as a function of $50\%$ condensation temperatures \citep{Lodders2003}. Refractory elements are on the left, volatiles on the right. Refractory lithophile elements are the green points at normalized abundance of $\sim 1.5 \;( \approx 1/0.675 )$. Refractory siderophiles are the black points at a normalized abundance of $\sim 3 \; (\approx 1/0.325 )$.}		
	\end{center}
\end{figure}

Based on our concordance estimates of the elemental abundances of the primitive mantle and core, we estimate the bulk elemental abundances of the Earth with their weighted sum 
( Eq. \ref{eq:5} ). 
The resultant bulk Earth composition and its associated uncertainty ( Eq. \ref{eq:6} ) are listed in column 7 of Table \ref{tab:earth}. 
The construction of bulk Earth abundances from primitive mantle and core for each element is plotted in Figs. \ref{fig:earth-1} and \ref{fig:earthTc} to demonstrate the geochemical differentiation between PM and core.
In Fig. \ref{fig:earthTc} we normalize the elemental abundances of PM and core to the concordance bulk Earth abundance and plot them as a function of 50\% condensation temperature \citep{Lodders2003}.

Our main results are shown in Fig. \ref{fig:earth-2} which shows the comparison of our concordance estimate of bulk Earth composition and the recent bulk Earth compositional datasets of \cite{Allegre2001}, \cite{McDonough2003}, and \cite{McDonough2008}. 
The light blue band in the lower panel indicates our estimate of the uncertainties on the concordance values. $70\%$ of the literature points fall within this band.
Fig. \ref{fig:earth-3} zooms in on the comparison of the 15 most abundant elements, which account for $99.90^{+0.10}_{-1.49}$ wt\% of the bulk Earth composition.

\section{Discussion}
\label{sec:disc}
\subsection{Comparison with Previous Estimates}
\label{sec:disc-1}

Compared to our concordance values, the abundances of some elements reported by \cite{McDonough2003} and \cite{McDonough2008} are significantly different (where the significance of the difference can only be based
on our reported uncertainties).
Relative to our values, their abundances for Mg, Cr, Br, Sn, Cd, B and Be are significantly higher, while their abundances for O, Na, Mn, Zn, Ga, Li, Rb, Nb, Ta and In are significantly lower, where ``significant" means their value is outside our estimate of the uncertainty (Figs. 6 \& 7).

The bulk elemental abundances with uncertainties in \cite{Allegre2001} were determined systematically by a carbonaceous chondrite correlation line, in combination with the abundances of siderophile and chalcophile elements in \cite{Allegre1995}. These works largely depend on the accuracy of the analyses of CI, CM, CO and CV meteorites compiled in \cite{Wasson1988}. For many elements, their mean values and/or their uncertainties are different from our results. For example,
relative to our values, the \cite{Allegre2001} abundances for O, Si, Mg, Sn, Rh, B and Cd are high, while their abundances for Fe, K, Cl, Zn, N, F, Ga, Tl, He and Ar are low.
One reason for these differences are the different assumptions made.
For example, the estimate of nitrogen abundance in the Earth in \cite{Allegre2001} is based on the atmospheric inventories while our estimate is based on the comparison of Earth's mantle with carbonaceous chondrite data in \cite{McDonough2008} and \cite{Palme2014b}. 
This latter assumption is more widely accepted and better constrained. 
\cite{Allegre2001} deduced Earth's chlorine abundance from the Cl/Ba ratio in MORB, while our PM sources \citep{Lyubetskaya2007, McDonough2008} derived Cl from the Cl/K ratio in MORB (Sect. 2.2). 
Other reasons for the abundance differences and differences in uncertainties are updates to PM and core abundances,
and the use of slightly different core mass fractions: \cite{McDonough2003} or \cite{McDonough2008} (32.3 wt\%), \cite{Allegre2001} (32.5 wt\%), and this work (32.5$\pm$0.3 wt\%).

Based on Eq. \ref{eq:7}, we quantify the significance of the deviation of our concordance bulk Earth abundances from previous estimates \citep{Allegre2001, McDonough2003, McDonough2008}. Among our new bulk elemental abundances, 6 elements (Mg, Sn, Br, B, Cd and Be) are more than $\sim 1\sigma$ below previous estimates, and 14 elements (Na, K, Cl, Zn, Sr, F, Ga, Rb, Nb, Gd, Ta, He, Ar, and Kr) are more than $\sim 1\sigma$ above previous estimates (see Table \ref{tab:S1}).

The reasons for these significant abundance differences include different assumptions and our inclusion of updated PM and core abundances. Mg and Na are the most abundant elements for which our new estimates deviate by more than $\sim 1.5\sigma$ from previous estimates. Our Mg PM abundance of 22.3$\pm$0.2 wt\% is $\sim 2.5\sigma$ lower than the 22.8 wt\% of \cite{McDonough2008} and $\sim 5\sigma$ lower than the 23.4 wt\% of \cite{Allegre2001} (deduced from their 15.8 wt\% Mg bulk abundance and a 67.5 \% PM mass fraction). 
Our lower bulk Earth Mg abundance predominantly comes from the lower PM Mg abundance in \cite{Palme2014b}.
Our bulk Earth Na abundance is higher than \cite{McDonough2008} and \cite{Allegre2001} because unlike those authors, we have included Na in the core:1372$\pm$519 ppm \citep{Kargel1993}.

\cite{Palme2014b} state that the 6 most abundant elements in the PM make up $98.41\pm0.01$ wt\%. 
We get a significantly higher value with a much larger uncertainty: $98.96 \pm 0.72$ wt\%. 
\cite{McDonough2016} states (based on chondritic models) that Fe, O, Si and Mg make up more than 90\% of the mass for the bulk Earth and the addition of Ni, Ca, Al and S accounts for more than 98\% by mass. Consistent with these estimates, we find $92.99\pm$ 1.45 wt\% for Fe, O, Si and Mg and 
$98.59^{+1.41}_{-1.47}$ wt\% with the addition of Ni, Ca, Al and S.    

\begin{figure}[H]
  \begin{center}
     \includegraphics[trim=0.9cm 2.2cm 0.5cm 3cm, scale=0.80,angle=0]{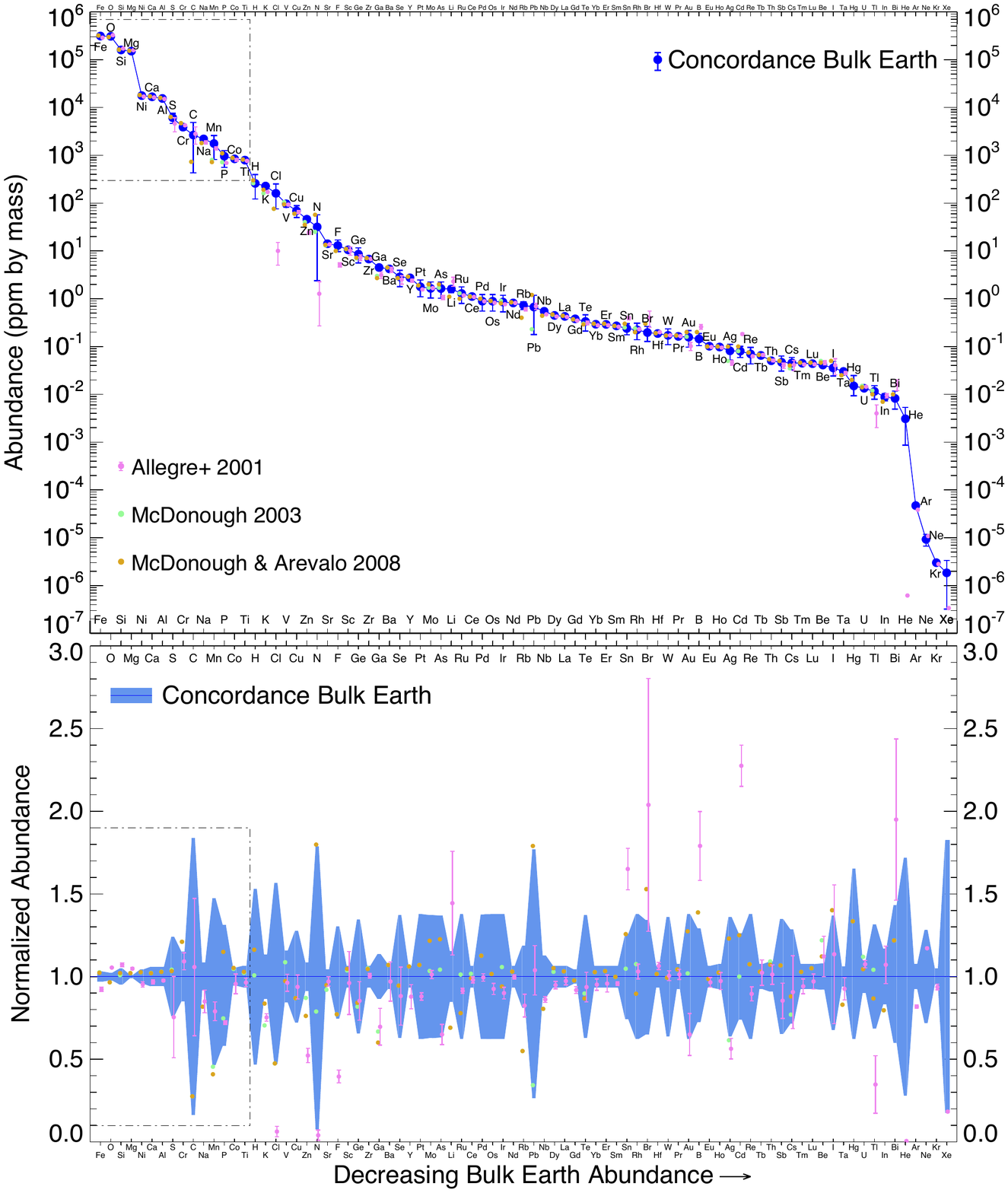} 
    \caption{Comparison of our concordance bulk Earth elemental abundances (last column of Table 1) with the recent estimates of \cite{Allegre2001}, \cite{McDonough2003}, and \cite{McDonough2008}. Elements for which \cite{McDonough2003} and \cite{McDonough2008} reported identical values are plotted as \cite{McDonough2008} points. In the lower panel the literature values from the top panel are normalized to our concordance values. The blue band in the lower panel indicates our estimate of the uncertainties on the concordance values. $70\%$ of the literature points fall within this band. The sum of our bulk Earth abundances is $10^6$ since we have rescaled the concordance PM and core abundances to $10^{6}$ (see Appendix C). The literature abundances have not been rescaled to ensure their abundances sum to $10^{6}$. The dashed boxes on the left in both panels contain the 15 most abundant elements and are zoomed-in on in Fig. \ref{fig:earth-3}.}
    \label{fig:earth-2}
  \end{center}
\end{figure}
\begin{figure}[H]
	\begin{center}
		\includegraphics[trim=0.9cm 2.2cm 0.5cm 3cm, scale=0.8,angle=0]{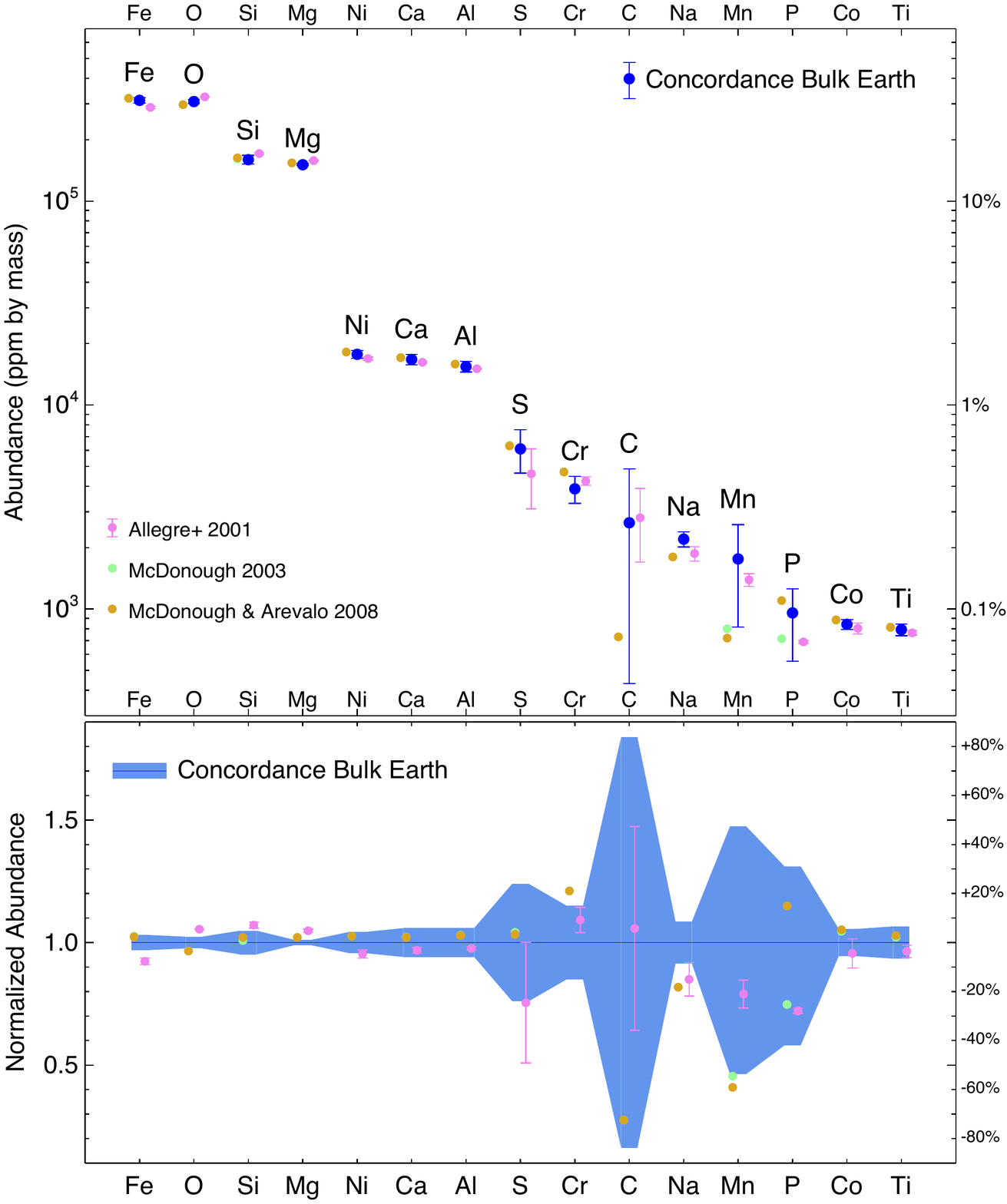} 
		\caption{Zoom-in of the 15 most abundant elements contained in the dashed boxes in both panels of the previous figure. The sum of the abundances of the 4 most abundant elements (Fe, O, Si, Mg) make up $92.99 \pm 1.45\%$ of the total mass of the Earth. The 8 most abundant elements (Fe, O, Si, Mg, Ni, Ca, Al, S) make up $98.59^{+1.41}_{-1.47} \%$, while the 15 most abundant elements plotted here make up $99.90^{+0.10}_{-1.49} \%$ of the total mass of the Earth. Our Mg abundance is significantly lower than previous estimates. Our estimates of the uncertainties allow the evaluation of the significance of such differences. Reported values without uncertainties do not allow such a comparison.}
		\label{fig:earth-3}
	\end{center}
\end{figure}
\begin{figure}[H]
	\begin{center}
		\includegraphics[trim=1.2cm 1.2cm 1.0cm 1.5cm, scale=0.6,angle=90]{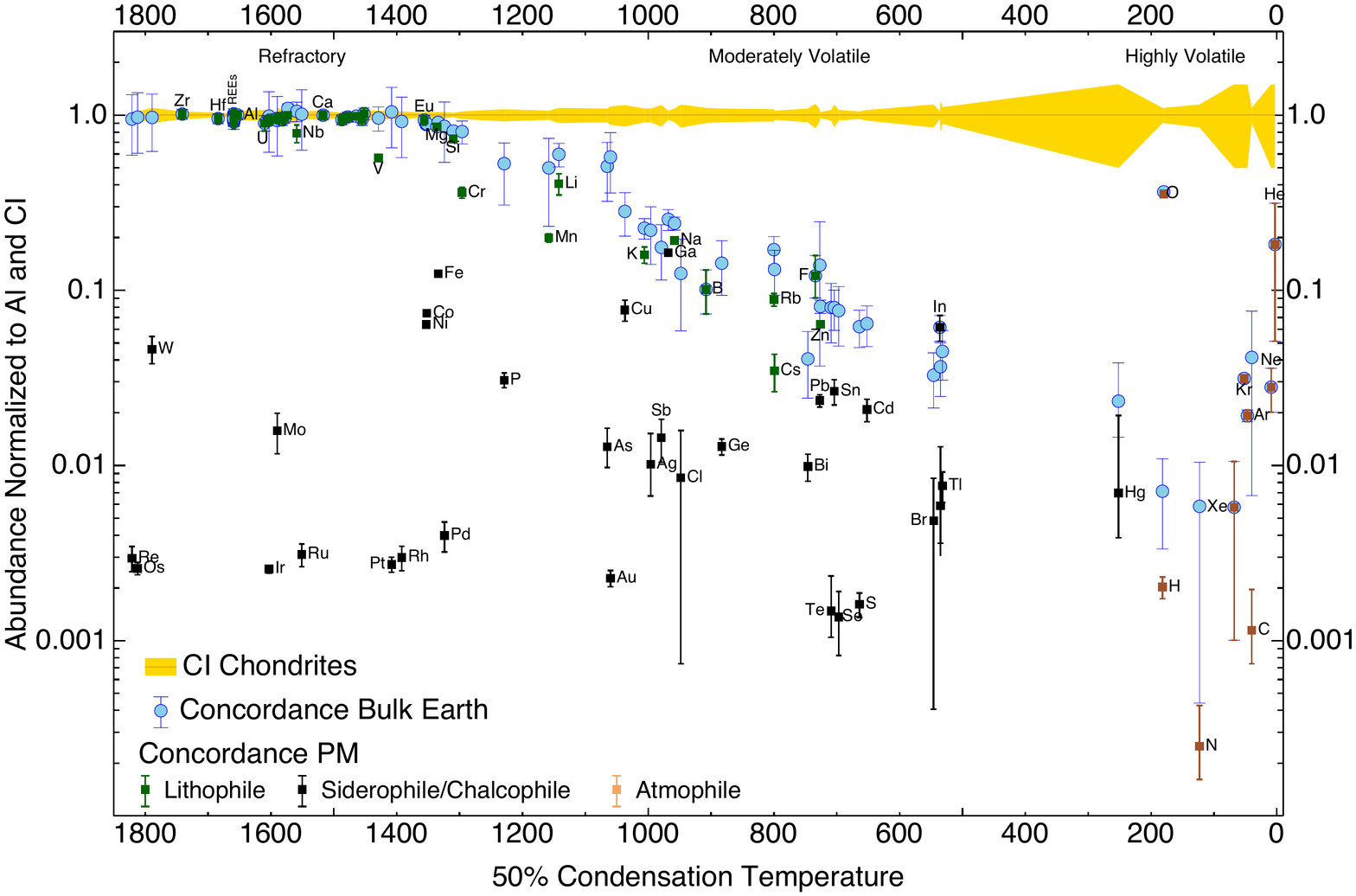} 
		\caption{Our concordance bulk Earth and primitive mantle abundances normalized to Al and CI chondrites. Thus, for a generic element {\it X}, the bulk Earth (blue dots) we plot: $(X/\textup{Al})_{Earth}/(X/\textup{Al})_{CI}$ and analogously for the concordance PM (square points) we plot: $(X/\textup{Al})_{PM}/(X/\textup{Al})_{CI}$. Both are plotted as a function of $50\%$ condensation temperatures \citep{Lodders2003}. CI abundances and uncertainties are from Table 3 of \cite{Palme2014a}, except for the uncertainties of the noble gases which we have set at the same $\pm50$\% uncertainty of Hg.}
		\label{fig:normCI}
	\end{center}
\end{figure}

Fig. \ref{fig:normCI} presents another way to present our PM and bulk Earth compositions (with uncertainties) in comparison with the latest compilation \citep{Palme2014a} of CI chondritic abundances.
In Fig. \ref{fig:normCI}, we have normalized both the bulk Earth and PM abundances to the highly refractory element Al and CI chondrites.
More specifically, for a generic elemental abundance {\it X}, for the bulk Earth we plot: $(X/\mathrm{Al})_{\mathrm{Earth}}/(X/\mathrm{Al})_{\mathrm{CI}}$ 
and for the concordance PM we plot: $(X/\mathrm{Al})_{\mathrm{PM}}/(X/\mathrm{Al})_{\mathrm{CI}}$.
This figure can be directly compared with Fig. 1 of \cite{Wood2006}, Fig. 1 of \cite{Carlson2014} and Fig. 21 of \cite{Palme2014b}.
Those three figures have been normalized to Mg and CI chondrites.
Because of their large abundances, Mg or Si have often been chosen as a normalization reference element \citep{McDonough2003, Palme2014b, Carlson2014, Litasov2016}.
However both Mg and Si are not strictly refractory elements. They both have condensation temperatures slightly lower than the transition or critical temperature ($\sim 1400$ K), below which the devolatilization of the Earth, compared to CI, is clear.
Thus, with an Mg normalization, a slight depletion of bulk Earth Mg compared to CI chondritic Mg, is misrepresented as an enrichment of refractory lithophiles.

Our PM analysis does not include estimates from \cite{Javoy2010}
who only report the PM abundances of ten elements (O, Mg, Si, Fe, Al, Ca, Ti, Ni, Cr, and Co) based on their enstatite chondrite model. If these abundances were included in our analysis they would lower Al, Ca and Ti without lowering the other RLEs such as Be, Sc, Sr, Nb, REE (Rare Earth Elements), Th and U. Also, enstatite chondrites are silica-enriched (but oxygen depleted?), from which it is problematic to construct the silica-poor (peridotitic) terrestrial model \citep{Fitoussi2012, Jellinek2015}. 

\subsection{Unresolved Issues}
\label{sec:disc-3}
We have assumed that the bulk Earth consists of the primitive mantle and the core, and that the primitive mantle is a reservoir 
with the composition of the present-day Earth's mantle, crust and surface inventories taken together. 
Contributions of late accretion to the formation of a secondary atmosphere of the Earth introduces some ambiguity to this definition.
We also neglected Earth's primary atmosphere (likely dominated by H and He) formed at the stage of solar nebula. 
In limiting our calculations to the prevalent standard models of the PM \citep{Lyubetskaya2007, McDonough2008, Palme2014b} we ignore the possible heterogeneity between the lower mantle and the upper mantle.
Impact erosion is another issue that could have changed the PM abundances of incompatible lithophile elements. If we considered the impact erosion model of \cite{ONeill2008}, for example, the PM abundance of K would decrease by about a factor of 2. 

Recently, based on i) a high PM abundance of the chalcophile element In, ii) nucleosynthetic isotope anomalies and iii) high- and low-pressure-temperature metal-silicate partitioning data, \cite{WangZ2016} suggested that Earth's moderately volatile element composition may not be chondritic.

Another concern could be that we compute weighted averages when combining the elemental abundances of the primitive mantle while we compute unweighted averages for the core. Considering the global consistency of the PM abundance datasets \citep[i.e.,][]{Lyubetskaya2007, McDonough2008, Palme2014b}, computing weighted averages is appropriate. 
Unlike the PM, core compositional models are much more complex and mixed with experimental data. Furthermore, uncertainties are not given in the majority of reported core compositional estimates, so computing weighted averages is not an option. 

The degree to which the bulk compositions of Venus and Mars are different from the Earth \citep{Morgan1980, Wanke1988} or broadly similar \mbox{\citep{Taylor2013, Kaib2015, Fitoussi2016}} is still unclear.

\section{Summary and Conclusions}
As the solar nebula condensed, evaporated and fractionated to form the early Earth \citep{Wood2006,Carlson2014}, the chemical composition of the bulk Earth was set.
From a heterogeneous set of literature values, we present the most complete lists of the elemental abundances {\it with uncertainties} of the primitive mantle (PM), the core and the bulk Earth (Table \ref{tab:earth}, Figs. \ref{fig:mantle-1}-\ref{fig:core-2},\ref{fig:earth-2},\ref{fig:earth-3}). The four most abundant elements (O, Mg, Si, and Fe) make up 94.19$\pm$0.69\% of the total PM mass. Fe-Ni alloy accounts for 87.90$\pm$2.92 wt\% of the total mass of the core, and the major light elements in the core are Si, O, S, C, Cr, Mn, P, Co, Na, Mg and H in order of decreasing abundance. 
The concordance bulk Earth abundances with uncertainties come from the weighted average of our concordance PM and core.
The weighting factor for this average comes from our new estimate (with uncertainty) of the core mass fraction of the Earth: 32.5$\pm$0.3 wt\%. 
Our concordance estimate of bulk Earth composition is largely consistent with recent bulk elemental abundance estimates; 
70\% of the previous bulk elemental abundances are within the uncertainties of our concordance bulk elemental abundances. 
Compared to previous work, the most significant differences include:  1) our abundances of Mg, Sn, Br, B, Cd and Be are more than $\sim 1 \sigma$ lower, and 2) our abundances of Na, K, Cl, Zn, Sr, F, Ga, Rb, Nb, Gd, Ta, He, Ar and Kr, more than $\sim 1 \sigma$ higher (Table B.4).
This set of concordance estimates (\textit{with uncertainties}) for the elemental abundances of PM, core and bulk Earth provides a reference that can 
be used to compare the Earth to the Sun, which will lead to a more precise devolatilization pattern, potentially applicable to exoplanets and their host stars.

\section*{Acknowledgments}
We thank two anonymous reviewers for insightful comments. We acknowledge useful discussions with William F. McDonough, Jun Korenaga, Hugh O'Neill, Hrvoje Tkal\v aci\'c, Brian Kennett and Marc Norman. 
H.S.W. acknowledges support from a Prime Minister's Australia Asia Endeavour Award and an Australian National University Postgraduate Research Scholarship.    

\appendix
\section{Concordance Estimates}
\label{appendix-1}
\subsection{Concordance PM Estimates}
\label{sec:method1}
For a given elemental abundance $X$, we compute the weighted mean $\overline{X}$ and variance $\sigma^2$ from:

\begin{equation}
	\label{eq:1}
	\overline{X} = \frac{\sum X_i/\sigma_i^2}{\sum 1/\sigma_i^2}
\end{equation}  

\begin{equation}
	\label{eq:2}
	\sigma^2 = \frac{1}{\sum 1/\sigma_i^2}
\end{equation}
where the index $i$ refers to a data set and ranges from $1$ to $N$.  
For most PM abundances $N=3$.
Since \cite{Lyubetskaya2007} did not report abundances for C, H, N and O, $N=2$ for these elements.
\cite{Lyubetskaya2007} abundances for Cl and Br are inconsistent with Cl and Br from the other two data sets.
Therefore, for these 2 elements, our concordance abundance is the unweighted mean (Eq. \ref{eq:3}) and 
we take as its uncertainty the range from the highest to the lowest reported abundance.

We have treated noble gases differently.
\cite{Marty2012} using an atmospheric model and \cite{Halliday2013} using three different models (layered mantle, impact erosion, and basaltic glass) report the molar abundances of non-radiogenic nuclides of noble gases  ($^3$He, $^{20}$Ne, $^{36}$Ar, $^{84}$Kr, $^{130}$Xe).
We convert these to atomic abundances (by number) by dividing by their estimated terrestrial primordial isotopic fractions \citep{Lodders2009}. 
We then convert to mass ppm using the atomic weights from \cite{Wieser2013}. 
Our concordance PM abundances for noble gases are the median value of the highest upper limit and the lowest lower limit among 
the four models.  We use the upper and lower limits as the uncertainty.

\subsection{Concordance Core Estimates}
\label{sec:method2}
Since the literature core abundances are more scattered and model-dependent than the PM abundances, we compute our concordance core abundances as unweighted means and 
assign the standard deviations as uncertainties using:

\begin{equation}
	\label{eq:3}
	\overline{X} = \frac{1}{N} \sum_{i=1}^{N}{X_i}
\end{equation}

\begin{equation}
	\label{eq:4}
	\sigma^2 = \frac{\sum{(X_i-\overline{X})^2}}{N-1}
\end{equation} 

Little work has been done on trace elements in the core. 
\cite{Kargel1993} and \cite{McDonough2008} provide estimates for many trace elements but their estimates are not independent. 
Thus Eq. \ref{eq:4} severely under-estimates the uncertainty on the abundances of these trace elements. 
To compensate for this underestimate, we do the following.
If the standard deviation from Eq. \ref{eq:4} is smaller than $\pm 40 \%$ we assign an uncertainty of $40\%$ (which is the average of the uncertainies of the 10 most abundant elements in the core).
If the standard deviation computed from Eq. \ref{eq:4} extends beyond the range of the reported values, we report the range of values as the uncertainty.
For those elements with only one reported value (i.e. N=1), Eq. \ref{eq:4} is undefined. For these cases, we report as an uncertainty either $\pm 40\%$, or the uncertainty on the single point, whichever is larger.

\subsection{Approach for concordance bulk Earth estimate}
\label{sec:method3}
The concordance elemental abundances of the bulk Earth for each element are computed by 
\begin{equation}
\label{eq:5}
X = X_{core} f_{core} + X_{PM}(1-f_{core})
\end{equation}  
where, $f_{core}$ is the core mass fraction estimated in Sect. \ref{sec:earth-1}. The uncertainty $\sigma_{X}$, associated with the bulk elemental abundance, is calculated by the error propagation of the three uncertainties: $\sigma_{X_{PM}}$, $\sigma_{X_{core}}$, and $\sigma_{f_{core}}$ of the PM elemental abundance, the core elemental abundance, and the core mass fraction, respectively:
\begin{equation}
\label{eq:6}
\sigma_{X}^2 = \sigma_{X_{core}}^2 f_{core}^2 + \sigma_{X_{PM}}^2 (1-f_{core})^2 +  \sigma_{f_{core}}^2 (X_{PM} - X_{core})^2
\end{equation}  
\section{Calculation of significance of deviation}
\label{appendix-2}
In Figs. 7 \& 8 we compare our concordance bulk estimates with three previous estimates of bulk Earth abundances.
For some elements, all three previous estimates are higher, or lower. For these elements we compute the
significance $S_X$, of this difference:
 
\begin{equation}
\label{eq:7}
S_X = \frac{1}{N} \sum_{i=1}^{N} \frac{X_{i} - X}{ \sqrt{\sigma_{X_{i}}^2 + \sigma_{X}^2}}
\end{equation}  
where, $X$ is our concordance estimate of the elemental abundance and $\sigma_{X}^2$ is its variance from Eq. \ref{eq:6}. 
$X_{i}$ is the abundance reported in the $i$-th literature source, with $N$ being the total number of literature sources. Identical abundances reported in \mbox{\cite{McDonough2003}} and in \mbox{\cite{McDonough2008}} have only been used once in this calculation.
$\sigma_{X_{i}}^2$ is its variance. If the source has no reported variance, we set $\sigma_{X_{i}}^2 = 0$.

Table \ref{tab:S1} lists the significances (more than $\sim 1 \sigma$) of the deviations between our concordance bulk Earth abundances and previous estimates \citep{Allegre2001, McDonough2003, McDonough2008}.
\begin{table*}
	{\small
		\caption{Significances$^a$ of deviations between our concordance bulk Earth abundances and previous estimates$^b$}
		\begin{center}
			\begin{tabular}{cccccccccccccc}
				\toprule
				\multicolumn{14}{l}{\bf 6 Elements with abundances more than $\sim 1 \sigma$ \textit {below} previous estimates} \\
				\hline
				Mg &Cd &B  &Be &Br &Sn & &&&&&&& \\
				-2.60 &-1.79 &-1.76 &-1.71 &-1.40 &-1.15 & &&&&&&&\\
				\hline\hline
				\multicolumn{14}{l}{\bf 14 Elements with abundances more than $\sim 1 \sigma$ \textit{above} previous estimates} \\
				\hline
				Zn &Ga &Ar &Ta &Na &K &Rb &F &He &Cl &Nb &Kr &Sr &Gd\\
				3.04 &2.36 &2.34 &1.82 &1.75 &1.75 &1.64 &1.63 &1.39 &1.38 &1.30 &1.20 &1.19 &1.11 \\		
				\bottomrule
				\multicolumn{14}{l}{$^a$\footnotesize calculated by Eq. B.1.} \\	
				\multicolumn{14}{l}{$^b$\footnotesize \cite{Allegre2001}, \cite{McDonough2003}, and \cite{McDonough2008}} \\			
				
			\end{tabular}
		\end{center}
		\label{tab:S1}
	}
\end{table*}

\section{Rescaling Data}
\label{sec:rescale}
Elemental abundances are usually reported in ppm by mass. Thus, when all elements are estimated, the sum of their abundances should equal $10^6$.
The abundances of all elements are usually not reported. We use this $\Sigma \: ppm = 10^6$ constraint to rescale our PM, core (and thus our bulk) abundances.
Our concordance PM abundances summed to $0.997 \times 10^6$ so we rescaled all of our PM abundances up by $0.3$\%.
Our concordance core abundances summed to $1.020 \times 10^6$ so we rescaled all of our core abundances down by $2.0$\%.
In Table C.5 we sum the literature abundances (column 3). 
In column 4 we supplement the sums from column 3 with our rescaled concordance abundances.
Deviations from $1.000 \times 10^6$ in column 4 indicate the level of inconsistency in the literature with the  $\Sigma \: ppm = 10^6$ constraint, 
even after the missing elements have been supplemented with our concordance abundances.

\begin{landscape}
\begin{table*}
{\footnotesize
	\centering
	\caption{Rescaling the sums of elemental abundances (ppm by mass) for Earth components}
	\begin{tabular}{l ll l }
	\toprule
	Sources & \multicolumn{1}{p{3.5cm}}{Number of reported abundances} & \multicolumn{1}{p{3.5cm}}{Sum of reported abundances} & \multicolumn{1}{p{3.5cm}}{Sum of reported abundances after the missing elements have been supplemented using our rescaled concordance abundances} \\ 
	\hline\hline
	\multicolumn{4}{c}{\bf a. Primitive mantle} \\
	\hline
	\cite{Lyubetskaya2007}	&	\multicolumn{1}{p{5cm}}{74 \scriptsize  (missing C, H, N, O, and noble gases)}	&	 0.556 $\times10^6$	&	0.998 $\times10^6$ \\  
	\cite{McDonough2008}	&	78 \scriptsize  ( missing noble gases)	&	 1.000 $\times10^6$	&	 	1.000 $\times10^6$ \\  
	\cite{Palme2014b}	&	78 \scriptsize  ( missing noble gases)	&	 1.001 $\times10^6$	&	 1.001 $\times10^6$ \\  
	\textbf {Concordance PM}	&	\boldmath $83$ 	&	\boldmath $0.997 \times10^6$ (pre-scaled)	&	 -	\\  
	\hline\hline
	\multicolumn{4}{c}{\bf b. Core} \\
	\cite{Kargel1993}	&	41	&	1.000 $\times10^6$	&	1.058 $\times10^6$ \\  
	\cite{Allegre1995, Allegre2001}	&	\multicolumn{1}{p{5cm}}{9 \scriptsize (O, Si, P, S, Cr, Mn, Fe, Co, Ni)}	&	0.998 $\times10^6$	&	1.009 $\times10^6$ \\  
	\cite{McDonough2003}	&	\multicolumn{1}{p{5cm}}{5 \scriptsize (O, Si, Mn, Fe, Ni)}	&	0.967 $\times10^6$	&	1.009 $\times10^6$ \\  
	\cite{Wood2006}	&	\multicolumn{1}{p{5cm}}{4 \scriptsize (H, C, Si, S)}	&	0.067 $\times10^6$	&		0.990 $\times10^6$ \\  
	\cite{McDonough2008}	&	40	&	1.003 $\times10^6$	&		1.005 $\times10^6$ \\  
	\cite{Javoy2010}	&	\multicolumn{1}{p{5cm}}{6 \scriptsize (O, Si, Cr, Fe, Co, Ni)}	&	1.003 $\times10^6$	&		1.038 $\times10^6$ \\  
	\cite{Huang2011}	&	1 \scriptsize (O)	&	0.005 $\times10^6$	&		0.978 $\times10^6$ 	\\  
	\cite{Rubie2011}	&	\multicolumn{1}{p{5.5cm}}{11 \scriptsize (O, Si, S, V, Cr, Fe, Co, Ni, Nb, Ta, W)}	&	1.007 $\times10^6$	&	1.024 $\times10^6$ \\  
	\cite{Zhang2012}	&	\multicolumn{1}{p{5cm}}{7 \scriptsize (H, C, N, O, Mg, Si, P)}	&	0.036 $\times10^6$	&		0.947 $\times10^6$  \\  
	\cite{Hirose2013}	&	3 \scriptsize (O, Si, and S)	&	0.105 $\times10^6$	&		1.010 $\times10^6$ \\  
	\cite{Wood2013}	&	1 \scriptsize (C)	&	0.010 $\times10^6$	&		1.002 $\times10^6$  \\  
	\cite{Siebert2013}	&	2 \scriptsize (O and Si)	&	0.069 $\times10^6$	&		0.992 $\times10^6$  \\  
	\cite{Badro2014}	&	\multicolumn{1}{p{5cm}}{4 \scriptsize (C, O, Si, S)}	&	0.057 $\times10^6$	&	0.955 $\times10^6$ \\  
	\cite{Nakajima2015}	&	1 \scriptsize (C)	&	0.011 $\times10^6$	&		1.002 $\times10^6$ 	\\  
	\cite{Litasov2016}	&	\multicolumn{1}{p{5cm}}{4 \scriptsize (C, O, Si, S)}	&	0.101 $\times10^6$	&		0.998 $\times10^6$ \\  
	\textbf {Concordance Core}	&	\textbf {49}	&	\boldmath $1.020\times10^6$ (pre-scaled)	&	-	\\  
	\hline\hline
	\multicolumn{4}{c}{\bf c. Bulk Earth} \\
	\hline
	\cite{Allegre2001}	&	\multicolumn{1}{p{3.5cm}}{81 \scriptsize  (missing H and Hg)}	&	 1.007$\times10^6$	&	 	1.007$\times10^6$  \\  
	\cite{McDonough2003}	&	78 \scriptsize  ( missing noble gases)	&	 1.002$\times10^6$	&		1.002$\times10^6$  \\  
	\cite{McDonough2008}	&	78 \scriptsize  ( missing noble gases)	&	 1.001$\times10^6$	&	 1.001$\times10^6$	\\  
	\textbf {Concordance Bulk Earth}	&	\boldmath $83$ 	&	\boldmath $1.000\times10^{6^*}$	&	 -	\\  
	\bottomrule
	\multicolumn{4}{p{11cm}}{\tiny $^*$ Based on rescaled concordance PM and concordance core.} \\	
	\end{tabular}
	\label{tab:S2}
}
\end{table*}
\end{landscape}


\section*{References}
\DeclareRobustCommand{\disambiguate}[3]{#1}
\bibliographystyle{elsarticle-harv} 
\bibliography{MyPapersBib}

\end{document}